\documentclass[10pt, conference, letterpaper]{IEEEtran}
\IEEEoverridecommandlockouts

\def\BibTeX{{\rm B\kern-.05em{\sc i\kern-.025em b}\kern-.08emT\kern-.1667em\lower.7ex\hbox{E}\kern-.125emX}}
    
\usepackage{url}
\usepackage{amsmath,amssymb,amsfonts}
\usepackage{algorithmic}
\usepackage{graphicx}
\usepackage{textcomp}
\usepackage{xcolor}
\usepackage{colortbl}
\usepackage{underscore}
\usepackage{multirow}
\usepackage{array}
\newcolumntype{H}{>{\setbox0=\hbox\bgroup}c<{\egroup}@{}}

\usepackage{booktabs}

\usepackage{subfigure}

\usepackage{fancyhdr}
\usepackage{extramarks}

\usepackage{rotating}
\usepackage[normalem]{ulem}
\newcommand{\revone}[2]{#2}

\newcommand{\revtwo}[2]{#2}

\newcommand{\revthree}[2]{#2}

\newcommand{\revfour}[2]{\textcolor{black}{#2}}

\newcommand{\commentout}[1]{}

\usepackage{todonotes}

\begin{document}

\title{Characterizing the Root Landscape of \\ Certificate Transparency Logs
\thanks{ISBN 978-3-903176-28-7 © 2020 IFIP}
}


\author{\IEEEauthorblockN{Nikita Korzhitskii}
\IEEEauthorblockA{
\textit{Linköping University, Sweden}\\
nikita.korzhitskii@liu.se} \and
\IEEEauthorblockN{Niklas Carlsson}
\IEEEauthorblockA{
\textit{Linköping University, Sweden}\\
niklas.carlsson@liu.se} 
}




\maketitle

\begin{abstract}

Internet security and privacy stand on the trustworthiness of public certificates signed by Certificate Authorities (CAs). However, software products do not trust the same CAs and therefore maintain different root stores, each typically containing hundreds of trusted roots capable of issuing ``trusted" certificates for any domain. 
Incidents with misissued certificates motivated Google to implement and enforce Certificate Transparency (CT).
CT logs archive certificates in a public, auditable and append-only manner.
The adoption of CT changed the trust landscape.
As a part of this change, CT logs started to maintain their own root lists and log certificates that chain back to one of the trusted roots.
\revfour{This has resulted in a yet-more complex trust ecosystem. In this paper,}{In this paper,}
we present the first characterization of this emerging CT root store landscape, as well as the tool that we developed for data collection, visualization, and analysis of the root 
\revfour{stores of the available CT logs.}{stores.} 
\revfour{As part of our characterization, w}{W}e compare the 
\revfour{}{logs'}
root stores and quantify their changes with respect to both each other and the root stores of major software vendors, look at evolving vendor CT policies, and show that root store mismanagement may be linked to log misbehavior.
\revthree{To gain further insights, we also surveyed the major log operators and vendors with CT log programs regarding their root store management and our observations.
Our results highlight ...}{Finally, we present and discuss the results of a survey that we have sent to 
\revfour{}{the}
log operators participating in Apple's and Google's CT log programs.}

\end{abstract}


\fancypagestyle{firststyle}
               {
                 \fancyhf{}
                 \fancyfoot[CF]{\tiny
                   \copyright IFIP (2020). This is the author's version of the work.
                   It is posted here by permission of IFIP for your personal use.
                   Not for redistribution. \\The definitive version was published in
                   {\em Proc. of IFIP Networking 2020},
                   Paris, France, June 2020,
       https://ieeexplore.ieee.org/document/9142756}
               }
               \thispagestyle{firststyle}

\section{Introduction}


The end-to-end security and privacy provided by HTTPS heavily depends on the trustworthiness of X.509 certificates and whether these certificates were signed by a Certificate Authority (CA) trusted by the client's software.  For example, most browsers 
\revthree{will establish}{establish} 
a connection using a public key from a certificate only if (i) it was directly signed by a CA included in their root store, or (ii) if it was signed by a sequence of intermediate CAs that chains back to one of these trusted roots.  However, not all CAs are equally trustworthy and 
software vendors 
\revfour{ended up using}{of browsers have decided to use} 
different root stores.
Furthermore, successful attacks, misconfigurations, and other incidents have significantly reduced the trust in CAs~\cite{leavitt2011internet}. 

Certificate Transparency (CT)~\cite{RFCCT} was 
\revfour{}{recently}
introduced as a response to waning CA 
\revfour{trust.}{trust. It provides an additional measure to improve the general trust in certificates, is already successfully deployed, and today plays a central and increasingly important role in the certificate ecosystem.}
With CT, certificates are logged in public,  auditable, append-only CT logs.  This helps domain owners to discover malicious or misissued certificates soon after they are issued.  
Certificates can be submitted to logs by any party. Upon submission logs issue Signed Certificate Timestamps (SCTs), which provide a cryptographic promise that the log will publish the submitted certificate chain within the log's Maximum Merge Delay (MMD). Servers can then deliver SCTs to their clients (e.g., browsers) to prove that a certificate has been logged. 

Both Google and Apple have implemented their own CT 
\revfour{policies, while CAs 
\revtwo{}{have}
started submitting their certificates to logs upon issuance.}{policies.}
For example, since 2015, 
\revfour{Google Chrome requires all newly issued Extended Validation (EV) certificates}{as part of Google Chrome's validation process, all newly issued Extended Validation (EV) certificates are required} 
to be logged in (at least) two Chrome-trusted CT logs (one Google-operated log and one non-Google-operated 
\revfour{log~\cite{ChromiumCTPolicy});}{log~\cite{ChromiumCTPolicy}), and}
since the release of 
Chrome v.68 in Jul. 2018, 
\revfour{{\it all} certificates issued after Apr. 2018 are required}{Chrome requires {\it all} certificates issued after Apr. 2018}
to be logged~\cite{ChromeFullCT}. 
Similarly, since Oct. 15, 2018, Apple requires all newly issued certificates to be included in several logs~\cite{CTPolicyApple}.
Neither Mozilla nor Microsoft have a public CT policy, 
but some Microsoft products 
\revthree{already have}{have} 
optional CT support~\cite{MozillaCT,MicrosoftCT}.

\revfour{}{To comply with the above policies, CAs (and sometimes domain owners) are actively submitting their certificates to logs upon issuance. Over recent years, the CT logs are therefore quickly growing in size.  For example, between January 2019 and January 2020, the number of entries in the CT logs tracked by Google increased from $\approx$3B entries to more than 7.5B entries~\cite{report}.}

By analogy with vendor root stores, there exist many CT logs and log operators configure their own lists of ``trusted" or ``acceptable"~\cite{RFCCT} root CAs.  
\revthree{Naturally, since logs only accept}{Since logs only accept} 
certificates 
\revthree{that chain back to these selected roots,}{chaining back to these roots,} 
\revthree{these 
roots}{the roots} 
determine 
\revthree{the portion of certificates that}{which certificates} 
can be 
\revfour{logged.}{logged (and indirectly also which certificates can be trusted by Chrome and Apple products).}  
\revthree{Furthermore, while the standard suggests that root lists ``might usefully be the union of root certificates trusted by major browser vendors''~\cite{RFCCT}, this does not happen in practice. Instead, as demonstrated in this paper, logs use diverse lists of acceptable roots, and most of them do not fully cover root stores of major vendors. Some logs also introduce additional custom submission requirements (e.g., expiry date within a certain range) that further restrict log selection.}{Some logs also introduce additional submission requirements (e.g., expiration constraints).  However, in general, it is important that certificates issued by trusted CAs can be easily submitted to multiple trusted logs.}



In this paper, we present a novel characterization of this emerging 
\revfour{trust}{root} 
landscape and the tool that we developed for data collection, visualization, and analysis of the logs' root stores.
First, we provide
an overview of the active CT logs, their root stores, 
\revthree{how the landscape and root stores are changing,}{describe the changing root store landscape,}
and 
\revthree{how they relate}{relate our observations} 
to current CT policies (by Google and Apple) and the major vendor root stores (by Apple, Microsoft, and Mozilla). We observed that root lists are diverse, 
\revthree{and}{that the root stores of most logs are increasing in size, becoming more similar, but also}
that most logs' root stores do not cover a significant fraction of the root stores of major software vendors, and that CT relies heavily on the logs operated by only 
\revthree{four log operators (Cloudflare, DigiCert, Google, and Sectigo).}{five log operators (Cloudflare, DigiCert, Google, Sectigo, and Let's Encrypt), where the logs of the fifth operator only became qualified by Google on Oct. 7, 2019.} 
This raises questions regarding whether the CT infrastructure is sufficiently redundant and reliable.
Second, we identify and highlight instances of 
\revfour{}{potential}
log mismanagement, 
including some instances that were followed by log misbehavior events resulting in logs being distrusted.
Third, we highlight a number of notable roots and specific log behaviors that demonstrate the diversity in the ways the logs are used.  
\revthree{}{The discussion is followed by the results of our survey of the five 
\revfour{primary log operators.}{log operators with
Apple- and Google-trusted CT logs.}}
Finally, along with this paper, we publish our interactive, online, open-source tool and 
\revthree{the}{a longitudinal} 
dataset to allow others to further investigate the root stores and their relations.
 
The paper is organized as follows. 
Section~\ref{sec:trusted} describes how Chrome and Apple decide which logs to trust.
Section~\ref{sec:tool} presents the tool and 
\revthree{the dataset.}{dataset.} 
Section~\ref{sec:roots} characterizes logs' root stores, their relations to each other and to the stores of major software 
\revthree{vendors.}{vendors, as well as how the landscape is changing.}
The following 
\revthree{two sections highlight}{section presents our survey results (Section~\ref{sec:survey}), highlights} 
log mismanagement and misbehavior (Section~\ref{sec:mismanagement}), as well as notable roots and use cases (Section~\ref{sec:root-standouts}).
Finally, we present related work (Section~\ref{sec:related}) and conclusions (Section~\ref{sec:conclusions}).

\section{Trusted vs Non-trusted Logs}\label{sec:trusted}

Both Google and Apple maintain their own CT policies and lists of trusted CT logs, with candidate logs only becoming trusted after demonstrating policy compliance over a test period.
To become Chrome-trusted (or ``qualified"), a log has to comply with the CT standard~\cite{RFCCT},
properly incorporate accepted certificate chains within the MMD, and satisfy availability constraints~\cite{ChromiumCTPolicy}. Moreover, the root store of every log must include Google's 
\revthree{{\it Merge Delay Monitor Root},}{{\it MMD Root},} 
which Google's monitoring servers use to perform test submissions. 
Logs are continually monitored, and Google disqualifies logs that violate policy requirements.
A qualified log must accept all certificates that chain back to the log's root store, however, it is allowed to reject submissions on the basis of validity, revocation, and/or expiration 
\revthree{status of a submitted certificate.}{status.}
Finally, Chrome-trusted logs have to publish their root lists and post updates in a corresponding thread of Google's CT bugtracker~\cite{bugtracker}. However, these announcements intertwine with other log-related posts and do not provide a solid, tamper-proof, and machine-readable history of 
\revfour{trust changes.}{changes in the root lists.}


Since Apple's CT policy~\cite{CTPolicyApple} has no public repository, it is more difficult to track changes over time.  
However, we have observed some updates to the policy, including a new requirement, stating that ``a log must accept certificates that are issued by Apple's compliance root CA to monitor the log's compliance with these policies"~\cite{AppleCTLogProgram}. 
Apple also states that ``logs must trust all root CA certificates included in Apple's trust store"~\cite{AppleCTLogProgram}. 
As we show later in the paper, logs do not fully adhere to these two requirements. \revthree{}{Both log programs adopt a similar log list schema; participating logs are assigned one of the following states: usable, qualified, read-only, retired, pending, or rejected.}



\section{Root Explorer and the Dataset}\label{sec:tool}

\begin{table*}[t]
\caption{Coverage of Vendor Root Stores and Other Properties of Available Certificate Transparency Logs during October 8th, 2019 as compared to December 27th, 2018}
\label{tab:logs2}
\vspace{-8pt}
\resizebox{\textwidth}{!}{%
\begin{tabular}{llllrlllccrlrlrlcccc}
\hline
\textbf{\textbf{Log}} & \textbf{\rotatebox{90}{\textbf{Distinct list}}} & \textbf{\rotatebox{90}{\textbf{\begin{tabular}{@{}l@{}}Google \\ Log Program\end{tabular}}}} & \textbf{\rotatebox{90}{\textbf{\begin{tabular}{@{}l@{}}Apple \\ Log Program\end{tabular}}}} & \multicolumn{1}{c}{\textbf{\rotatebox{90}{\textbf{\begin{tabular}{@{}l@{}}Distinct \\ certificates\end{tabular}}}}} & \textbf{\textbf{}} & \textbf{\rotatebox{90}{\textbf{Duplicates}}} & \textbf{\textbf{}} & \textbf{\rotatebox{90}{\textbf{\begin{tabular}{@{}l@{}}Merge Delay \\ Monitor Root\end{tabular}}}} & \textbf{\rotatebox{90}{\textbf{\begin{tabular}{@{}l@{}}DigiNotar \\ Root CA\end{tabular}}}} & \textbf{\rotatebox{90}{\textbf{Apple, \%}}} & \textbf{\textbf{}} & \textbf{\rotatebox{90}{\textbf{Microsoft, \%}}} & \textbf{\textbf{}} & \textbf{\rotatebox{90}{\textbf{Mozilla, \%}}} & \textbf{\textbf{}} & \textbf{\rotatebox{90}{\textbf{CORS headers}}} & \textbf{\rotatebox{90}{\textbf{Test submission}}} & \rotatebox{90}{\textbf{\begin{tabular}{@{}l@{}}Expiration \\ constraint\end{tabular}}} & \rotatebox{90}{\textbf{Rejects expired}} \\ \hline
Cloudflare Cirrus (RPKI log) & 1 & \cellcolor[HTML]{EFEFEF}not listed & \cellcolor[HTML]{EFEFEF}not listed & 5 &  &  &  & \cellcolor[HTML]{EFEFEF}- &  & 0.0 &  & 0.0 &  & 0.0 &  & \cellcolor[HTML]{FFFFC7}- & + & \cellcolor[HTML]{EFEFEF} & \cellcolor[HTML]{EFEFEF} \\ \hline
Cloudflare Nimbus20\{17-18\} & 2 & \cellcolor[HTML]{FD6864}rejected & \cellcolor[HTML]{FFFFC7}readonly & 576 & {\color[HTML]{32CB00} +213} &  &  & + & \cellcolor[HTML]{FFFC9E}+ & \cellcolor[HTML]{9AFF99}100 & {\color[HTML]{32CB00} +9.8} & 97.5 & {\color[HTML]{32CB00} +29.7} & \cellcolor[HTML]{9AFF99}100 & {\color[HTML]{32CB00} +16.7} & \cellcolor[HTML]{FFFFC7}- & \cellcolor[HTML]{FD6864}– & + & \cellcolor[HTML]{FFCC67}+ \\ \hline
Cloudflare Nimbus20\{19-21\} & 2 & \cellcolor[HTML]{67FD9A}usable & \cellcolor[HTML]{67FD9A}usable & 576 & {\color[HTML]{32CB00} +213} &  &  & + & \cellcolor[HTML]{FFFC9E}+ & \cellcolor[HTML]{9AFF99}100 & {\color[HTML]{32CB00} +9.8} & 97.5 & {\color[HTML]{32CB00} +29.7} & \cellcolor[HTML]{9AFF99}100 & {\color[HTML]{32CB00} +16.7} & \cellcolor[HTML]{FFFFC7}- & + & + & – \\ \hline
Cloudflare Nimbus20\{22-23\} & 2 & \cellcolor[HTML]{9AFF99}qualified & \cellcolor[HTML]{67FD9A}usable & 576 & {\color[HTML]{32CB00} +213} &  &  & + & \cellcolor[HTML]{FFFC9E}+ & \cellcolor[HTML]{9AFF99}100 & {\color[HTML]{32CB00} +9.8} & 97.5 & {\color[HTML]{32CB00} +29.7} & \cellcolor[HTML]{9AFF99}100 & {\color[HTML]{32CB00} +16.7} & \cellcolor[HTML]{FFFFC7}- & + & + & \cellcolor[HTML]{EFEFEF} \\ \hline
DigiCert Log Server & 3 & \cellcolor[HTML]{67FD9A}usable & \cellcolor[HTML]{67FD9A}usable & 56 & {\color[HTML]{32CB00} +4} & \multicolumn{1}{r}{\cellcolor[HTML]{FFFC9E}1} & {\color[HTML]{CB0000} +1} & + &  & 27.5 & {\color[HTML]{32CB00} +1.6} & 15.7 & {\color[HTML]{32CB00} +3.1} & 32.9 & {\color[HTML]{32CB00} +2.9} & \cellcolor[HTML]{FFFFC7}- & + & – & – \\ \hline
DigiCert Log Server 2 & 4 & \cellcolor[HTML]{67FD9A}usable & \cellcolor[HTML]{67FD9A}usable & 179 & {\color[HTML]{32CB00} +3} & \multicolumn{1}{r}{\cellcolor[HTML]{FFFC9E}1} &  & + &  & 79.2 & {\color[HTML]{CB0000} -3.0} & 45.8 & {\color[HTML]{32CB00} +3.7} & 87.9 & {\color[HTML]{CB0000} -2.8} & \cellcolor[HTML]{FFFFC7}- & \cellcolor[HTML]{FFCC67}± & – & – \\ \hline
DigiCert Nessie2018 & 5 & \cellcolor[HTML]{FD6864}rejected & \cellcolor[HTML]{FD6864}rejected & 531 & {\color[HTML]{32CB00} +6} & \multicolumn{1}{r}{\cellcolor[HTML]{FFFC9E}2} & {\color[HTML]{CB0000} +1} & + &  & 97.2 & {\color[HTML]{CB0000} -2.2} & 87.1 & {\color[HTML]{CB0000} -4.0} & 93.3 & {\color[HTML]{CB0000} -2.0} & \cellcolor[HTML]{FFFFC7}- & \cellcolor[HTML]{FD6864}– & + & \cellcolor[HTML]{FFCC67}+ \\ \hline
DigiCert Yeti2018 & 5 & \cellcolor[HTML]{FD6864}rejected & \cellcolor[HTML]{67FD9A}usable & 531 & {\color[HTML]{32CB00} +6} & \multicolumn{1}{r}{\cellcolor[HTML]{FFFC9E}2} & {\color[HTML]{CB0000} +1} & + &  & 97.2 & {\color[HTML]{CB0000} -2.2} & 87.1 & {\color[HTML]{CB0000} -4.0} & 93.3 & {\color[HTML]{CB0000} -2.0} & \cellcolor[HTML]{FFFFC7}- & \cellcolor[HTML]{FD6864}– & + & \cellcolor[HTML]{FFCC67}+ \\ \hline
DigiCert Nessie20\{19-22\}/Yeti20\{19-22\} & 5 & \cellcolor[HTML]{67FD9A}usable & \cellcolor[HTML]{67FD9A}usable & 531 & {\color[HTML]{32CB00} +6} & \multicolumn{1}{r}{\cellcolor[HTML]{FFFC9E}2} & {\color[HTML]{CB0000} +1} & + &  & 97.2 & {\color[HTML]{CB0000} -2.2} & 87.1 & {\color[HTML]{CB0000} -4.0} & 93.3 & {\color[HTML]{CB0000} -2.0} & \cellcolor[HTML]{FFFFC7}- & + & + & \cellcolor[HTML]{FFCC67}+ \\ \hline
DigiCert Nessie2023/Yeti2023 & 5 & \cellcolor[HTML]{9AFF99}qualified & \cellcolor[HTML]{9AFF99}qualified & 531 &  & \multicolumn{1}{r}{\cellcolor[HTML]{FFFC9E}2} &  & + &  & 97.2 &  & 87.1 &  & 93.3 &  & \cellcolor[HTML]{FFFFC7}- & + & + & \cellcolor[HTML]{EFEFEF} \\ \hline
Google Argon2022 & 6 & \cellcolor[HTML]{9AFF99}qualified & \cellcolor[HTML]{67FD9A}usable & 561 & {\color[HTML]{32CB00} +24} &  &  & + &  & 99.4 & {\color[HTML]{32CB00} +0.0} & 97.2 & {\color[HTML]{32CB00} +2.4} & \cellcolor[HTML]{9AFF99}100 & {\color[HTML]{32CB00} +0.7} &  & + & + & \cellcolor[HTML]{EFEFEF} \\ \hline
Google Daedalus & 6 & \cellcolor[HTML]{EFEFEF}no state & \cellcolor[HTML]{EFEFEF}not listed & 561 & {\color[HTML]{32CB00} +24} &  &  & + &  & 99.4 & {\color[HTML]{32CB00} +0.0} & 97.2 & {\color[HTML]{32CB00} +2.4} & \cellcolor[HTML]{9AFF99}100 & {\color[HTML]{32CB00} +0.7} &  & + & + & – \\ \hline
Google Icarus & 7 & \cellcolor[HTML]{67FD9A}usable & \cellcolor[HTML]{67FD9A}usable & 3 &  &  &  & + &  & 1.1 & {\color[HTML]{32CB00} +0.0} & 0.6 & {\color[HTML]{32CB00} +0.1} & 1.3 & {\color[HTML]{32CB00} +0.0} &  & + & – & – \\ \hline
Google Skydiver & 8 & \cellcolor[HTML]{67FD9A}usable & \cellcolor[HTML]{67FD9A}usable & 559 & {\color[HTML]{32CB00} +24} &  &  & + &  & 98.3 & {\color[HTML]{32CB00} +0.0} & 96.6 & {\color[HTML]{32CB00} +2.4} & 98.7 & {\color[HTML]{32CB00} +0.7} &  & + & – & – \\ \hline
Google Solera2023 & 9 & \cellcolor[HTML]{EFEFEF}no state & \cellcolor[HTML]{EFEFEF}not listed & 225 &  &  &  & + &  & 1.1 &  & 0.6 &  & 1.3 &  &  & + & + & \cellcolor[HTML]{EFEFEF} \\ \hline
Google Submariner & 10 & \cellcolor[HTML]{EFEFEF}no state & \cellcolor[HTML]{EFEFEF}not listed & 81 & {\color[HTML]{32CB00} +1} &  &  & + &  & 20.8 & {\color[HTML]{CB0000} -1.6} & 20.3 & {\color[HTML]{32CB00} +3.0} & 15.4 & {\color[HTML]{32CB00} +0.1} &  & + & – & – \\ \hline
Google Crucible/Solera20\{18-22\}/Testtube & 9 & \cellcolor[HTML]{EFEFEF}no state & \cellcolor[HTML]{EFEFEF}not listed & 225 & {\color[HTML]{32CB00} +31} &  &  & + &  & 1.1 & {\color[HTML]{32CB00} +0.0} & 0.6 & {\color[HTML]{32CB00} +0.1} & 1.3 & {\color[HTML]{32CB00} +0.0} &  & + & \cellcolor[HTML]{EFEFEF} & \cellcolor[HTML]{EFEFEF} \\ \hline
Google Argon2017 & 6 & \cellcolor[HTML]{FD6864}rejected & \cellcolor[HTML]{67FD9A}usable & 561 & {\color[HTML]{32CB00} +24} &  &  & + &  & 99.4 & {\color[HTML]{32CB00} +0.0} & 97.2 & {\color[HTML]{32CB00} +2.4} & \cellcolor[HTML]{9AFF99}100 & {\color[HTML]{32CB00} +0.7} &  & \cellcolor[HTML]{FD6864}– & + & \cellcolor[HTML]{FFCC67}+ \\ \hline
Google Argon2018/Xenon2018 & 6 & \cellcolor[HTML]{FD6864}rejected & \cellcolor[HTML]{67FD9A}usable & 561 & {\color[HTML]{32CB00} +24} &  &  & + & \multicolumn{1}{l}{} & 99.4 & {\color[HTML]{32CB00} +0.0} & 97.2 & {\color[HTML]{32CB00} +2.4} & \cellcolor[HTML]{9AFF99}100 & {\color[HTML]{32CB00} +0.7} & \multicolumn{1}{l}{} & + & + & – \\ \hline
Google Argon20\{19-21\}/Xenon20\{19-22\} & 6 & \cellcolor[HTML]{67FD9A}usable & \cellcolor[HTML]{67FD9A}usable & 561 & {\color[HTML]{32CB00} +24} &  &  & + &  & 99.4 & {\color[HTML]{32CB00} +0.0} & 97.2 & {\color[HTML]{32CB00} +2.4} & \cellcolor[HTML]{9AFF99}100 & {\color[HTML]{32CB00} +0.7} &  & + & + & – \\ \hline
Google Pilot/Rocketeer & 6 & \cellcolor[HTML]{67FD9A}usable & \cellcolor[HTML]{67FD9A}usable & 561 & {\color[HTML]{32CB00} +24} &  &  & + &  & 99.4 & {\color[HTML]{32CB00} +0.0} & 97.2 & {\color[HTML]{32CB00} +2.4} & \cellcolor[HTML]{9AFF99}100 & {\color[HTML]{32CB00} +0.7} &  & + & – & – \\ \hline
Google Argon2023/Xenon2023 & 6 & \cellcolor[HTML]{9AFF99}qualified & \cellcolor[HTML]{9AFF99}qualified & 561 &  &  &  & + &  & 99.4 &  & 97.2 &  & \cellcolor[HTML]{9AFF99}100 &  &  & + & + & \cellcolor[HTML]{EFEFEF} \\ \hline
Let{\textsc{\char13}}s Encrypt Oak20\{19-22\} & 11 & \cellcolor[HTML]{9AFF99}qualified & \cellcolor[HTML]{9AFF99}qualified & 412 &  &  &  & + &  & \cellcolor[HTML]{9AFF99}100 &  & 99.4 &  & \cellcolor[HTML]{9AFF99}100 &  &  & + & + & \cellcolor[HTML]{FFCC67}+ \\ \hline
Nordu Plausible & 12 & \cellcolor[HTML]{EFEFEF}no state & \cellcolor[HTML]{EFEFEF}not listed & 444 &  &  &  & + &  & 81.5 & {\color[HTML]{CB0000} -4.7} & 68.3 & {\color[HTML]{CB0000} -6.6} & 76.5 & {\color[HTML]{CB0000} -2.8} & \cellcolor[HTML]{FFFFC7}- & + & – & – \\ \hline
Sectigo Dodo & 13 & \cellcolor[HTML]{EFEFEF}no state & \cellcolor[HTML]{EFEFEF}not listed & 522 & {\color[HTML]{32CB00} +73} &  &  & + &  & \cellcolor[HTML]{9AFF99}100 &  & \cellcolor[HTML]{9AFF99}100 & {\color[HTML]{32CB00} +0.3} & \cellcolor[HTML]{9AFF99}100 & {\color[HTML]{32CB00} +0.7} & \cellcolor[HTML]{FFFFC7}- & + & – & – \\ \hline
Sectigo Mammoth/Sabre & 14 & \cellcolor[HTML]{67FD9A}usable & \cellcolor[HTML]{67FD9A}usable & 371 & {\color[HTML]{32CB00} +14} &  &  & + &  & \cellcolor[HTML]{9AFF99}100 &  & 89.8 & {\color[HTML]{32CB00} +0.8} & 98.7 & -0.6 & \cellcolor[HTML]{FFFFC7}- & + & – & – \\ \hline
WoTrus & 15 & \cellcolor[HTML]{EFEFEF}not listed & \cellcolor[HTML]{EFEFEF}not listed & 9 &  &  &  & + &  & 1.7 &  & 0.0 &  & 0.0 &  & \cellcolor[HTML]{FFFFC7}- & \cellcolor[HTML]{FD6864}– &  & \cellcolor[HTML]{EFEFEF} \\ \hline \hline
CNNIC CT log &  & \cellcolor[HTML]{FFFC9E}retired & \cellcolor[HTML]{FFFC9E}retired & \multicolumn{1}{c}{\cellcolor[HTML]{FFCCC9}–} &  & \multicolumn{1}{r}{} &  &  &  &  &  &  &  &  &  &  & \cellcolor[HTML]{FD6864}– &  &  \\ \hline
GDCA Log 1/Log 2 &  & \cellcolor[HTML]{FD6864}rejected & \cellcolor[HTML]{FD6864}rejected & \multicolumn{1}{c}{\cellcolor[HTML]{FFCCC9}–} &  & \cellcolor[HTML]{FFFC9E}2 &  &  &  &  &  &  &  &  &  &  & \cellcolor[HTML]{FD6864}– &  &  \\ \hline
GDCA CT log \#1/SHECA CT log 2 &  & \cellcolor[HTML]{FD6864}rejected & \cellcolor[HTML]{EFEFEF}not listed & \multicolumn{1}{c}{\cellcolor[HTML]{FFCCC9}–} &  &  &  &  &  &  &  &  &  &  &  &  & \cellcolor[HTML]{FD6864}– &  &  \\ \hline
Google Aviator &  & \cellcolor[HTML]{FFFFC7}readonly & \cellcolor[HTML]{FFFFC7}readonly & \multicolumn{1}{c}{\cellcolor[HTML]{FFCCC9}–} &  & \multicolumn{1}{r}{} &  &  &  &  &  &  &  &  &  &  & \cellcolor[HTML]{FD6864}– &  &  \\ \hline
Venafi Gen2 CT log &  & \cellcolor[HTML]{FFFFC7}readonly & \cellcolor[HTML]{FFFFC7}readonly & \multicolumn{1}{c}{\cellcolor[HTML]{FFCCC9}–}&  & \cellcolor[HTML]{FFFC9E}6 &  &  &  &  &  &  &  &  &  &  & \cellcolor[HTML]{FD6864}– &  &  \\ \hline \hline
Apple Trust Store version 2018071800 &  &  &  & 174 &  &  &  &  &  & 94.9 &  & 46.2 &  & 81.2 &  &  &  &  &  \\ \hline
Apple Trust Store version 2018121000 &  &  &  & 178 &  &  &  &  &  & 100 &  & 47.4 &  & 85.2 &  &  &  &  &  \\ \hline
Microsoft Trusted Root Program (2018-10) &  &  &  & 382 &  &  &  &  &  & 91.6 &  & 96.6 &  & 97.3 &  &  &  &  &  \\ \hline
Microsoft Trusted Root Program (2019-07) &  &  &  & 325 &  &  &  &  &  & 86.5 &  & 100 &  & 97.3 &  &  &  &  &  \\ \hline
Mozilla Included CA Certificate List (2018-12) &  &  &  & 150 &  &  &  &  &  & 74.2 &  & 44.6 &  & 96.6 &  &  &  &  &  \\ \hline
Mozilla Included CA Certificate List (2019-10) &  &  &  & 149 &  &  &  & \multicolumn{1}{l}{} & \multicolumn{1}{l}{} & \multicolumn{1}{l}{71.3} &  & \multicolumn{1}{l}{44.6} &  & \multicolumn{1}{l}{100} &  & \multicolumn{1}{l}{} & \multicolumn{1}{l}{} & \multicolumn{1}{l}{} & \multicolumn{1}{l}{} \\ \hline
\end{tabular}%
}
\vspace{-0pt} \\
{\scriptsize Positive and negative colored values designate change between the initial and final measurements. ``+"/``-" in {\it Test submission} column designate whether the submission was successful.
The symbol ``±" indicates that only some submissions were successful.}
\end{table*}

To allow readers to dive deeper into our results and the CT landscape, we 
\revthree{}{will} 
share both our tool and the datasets~\cite{explorer}.

\begin{figure}[t]
  \centering
  \vspace{-0mm}
  \subfigure[Euler mode]{
  \includegraphics[trim = 0mm 0mm 0mm 00mm, clip, width=0.4\textwidth]{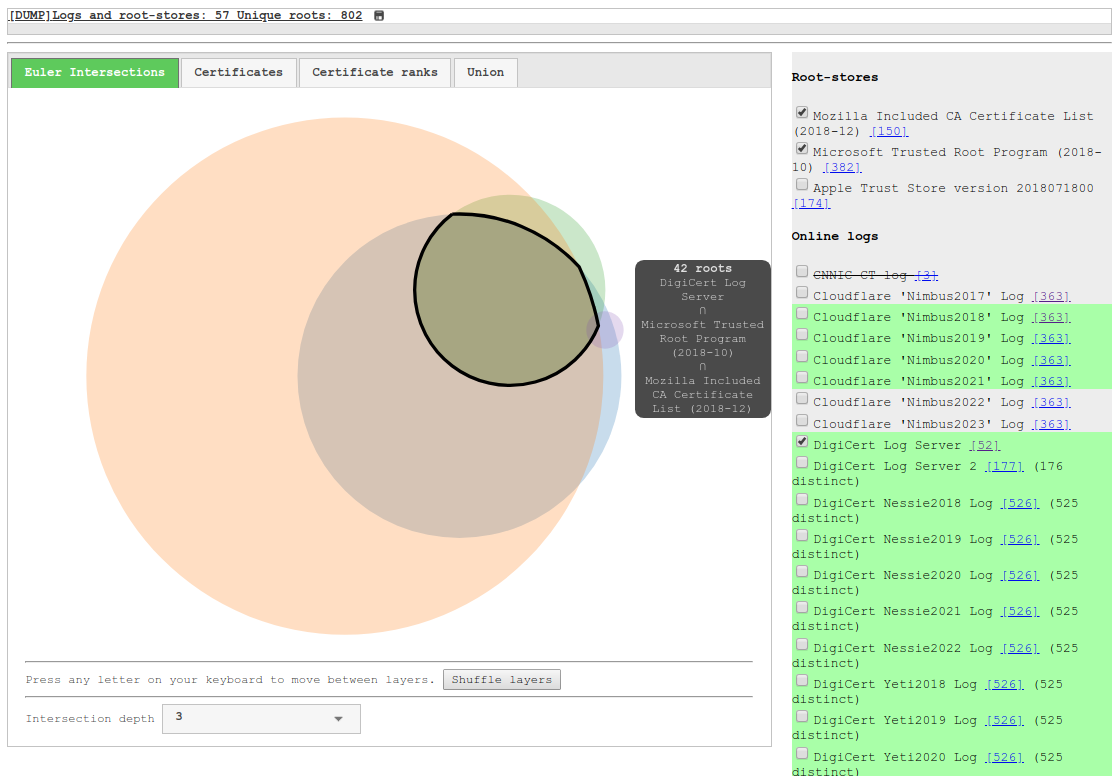}}
  \hspace{30pt}
   \subfigure[Listing mode]{
   \includegraphics[trim = 5mm 45mm 2mm 5mm, clip, width=0.4\textwidth]{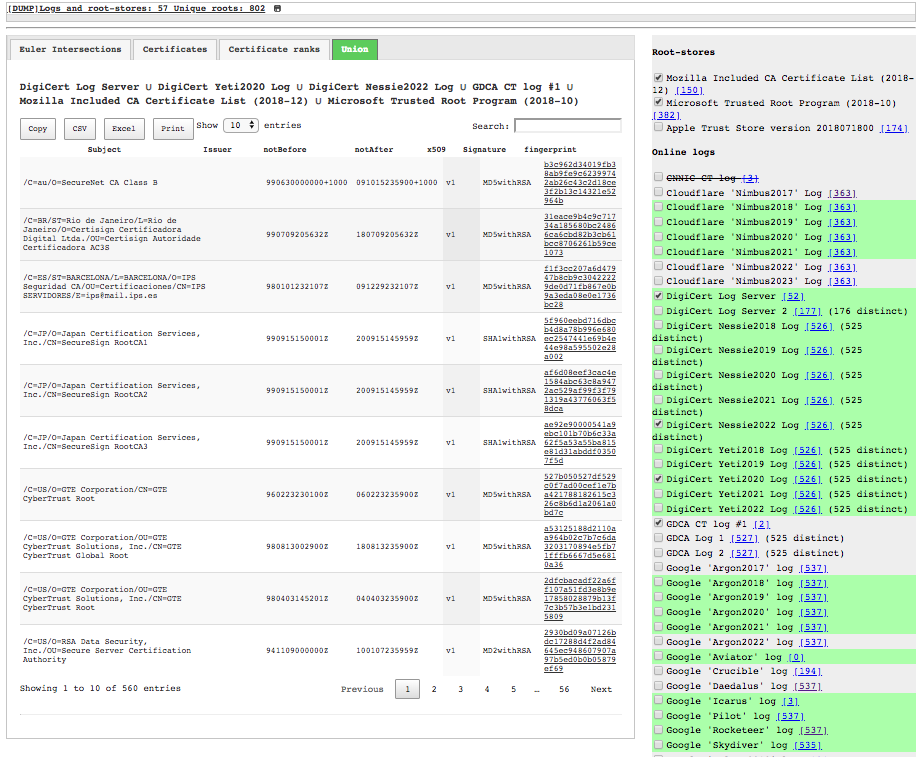}}
  \vspace{0pt}
  \caption{{\em Certificate Transparency Root Explorer} screenshots: (a) interactively selected intersection in Euler Diagram visualization mode, and (b) certificate-listing mode.}
    \label{fig:screenshots}
     \vspace{-8pt}
\end{figure}

{\bf Tool:} We developed {\em CT Root Explorer} --- a web-based open-source interactive tool for data collection and analysis of logs' root stores, including the relations to each other and to major vendor root stores. The tool can be used both on live data and historic data. The tool first retrieves roots directly from available logs and/or import/export SQLite snapshots.
The tool allows a user to interactively visualize and analyze selected root stores, 
explore certificate frequencies, intersections, complements, and unions. Information about roots can also be filtered and exported. Figure~\ref{fig:screenshots} shows the tool's interface in {\it Euler diagram} mode, in which a user can interactively select an intersection between a number of logs, and in {\it Certificate listing} mode, in which the information about a selected set of roots is listed and can be filtered and exported.  The tool also has two other modes, including the {\it Root frequency} mode, which generates an interactive frequency diagram (an annotated version is presented in Figure~\ref{fig:frequency}).





\revthree{{\bf Dataset:}}{{\bf Primary dataset:}}
\revthree{The default dataset was collected on Dec. 27, 2018, using {\em CT Root Explorer} running on Chromium v.71, and contains the root stores from all known and trusted logs listed by Google~\cite{LogsKnown, LogsChrome} and Apple~\cite{AppleCTLogProgram}  (54 in total).}{Since Nov. 2, 2018, we have collected hourly snapshots of the root stores of all logs listed as known by Google~\cite{LogsKnown,LogsChrome} and Apple~\cite{AppleCTLogProgram} using {\em CT Root Explorer} running on Chromium v.71.  The default snapshots used in the analysis presented here (and made available with the tool) were collected on Dec. 27, 2018, and Oct. 8, 2019.  These snapshots include the root stores of 54 and 57 logs, respectively.}
\revthree{We also collected and included the roots from the three popular vendor root stores by}{We also include corresponding snapshots of the three popular vendor root stores by}
Apple~\cite{AppleRoots, AppleRoots2}, Microsoft~\cite{MicrosoftRoots}, and Mozilla~\cite{MozillaRoots}. (Unlike Firefox, 
Google Chrome relies on the root store of the underlying operating 
system, but reserves the right to distrust selected CAs~\cite{ChromiumRootPolicy}.) 
Table~\ref{tab:logs2}
lists the collected logs and root stores. Here, some logs with identical properties are grouped by a year range 
\revfour{(e.g., Argon20\{18-20\}).}{(e.g., Argon20\{18-20\}), where the year typically reflects an additional expiration constraint on the certificates to be logged.}
\revthree{In the table, we}{We} 
also include {\it Cloudflare Cirrus} and {\it WoTrus} 
\revthree{\revthree{(designated with * in Table~\ref{tab:logs2}),}{(denoted an * in Table~\ref{tab:logs2}),}}{(``not listed"),} 
which are not listed as {\it known} by Apple or 
\revthree{Google, and therefore not included in the dataset itself and the other figures.}{Google.}
\revthree{}{For each log, we specify their current status in Google's and Apple's log programs.  Here, we distinguish between trusted logs (green) that are (i) usable, and (ii) qualified (i.e., accepted, but not yet used). Non-trusted logs include logs that are (i) rejected, (ii) not listed, or have (iii) no state.  At the boundary of this class (yellow) we include (iv) retired (previously usable/qualified) and (v) read-only (archived).}

{\bf Collection and basic log properties:}
During a live scan, the tool attempts to retrieve 
\revfour{trusted roots}{the logs' acceptable/trusted roots} 
using the {\it get-roots} method of each available log. However, since only Google's 
\revthree{}{and Let's Encrypt's}
logs include Cross-origin Resource Sharing (CORS) related headers in their responses (see ``CORS header" in Table~\ref{tab:logs2}), we had to disable the CORS enforcement in our browser (used to protect modern browsers)  to retrieve data from the other logs.  Furthermore, due to some logs having invalid server certificates 
\revthree{(see ``Log TLS conn. valid." in Table~\ref{tab:logs2})(CNNIC, SHECA and GDCA logs; all of which had certificates not trusted with respect to Apple and/or Mozilla root stores, and that are offline or have an expired certificate)}{} we disabled TLS certificate verification. 
\revthree{Both these
adjustments show that many 
\revtwo{of the logs}{logs} 
need to update their servers or certificates.  
These}{These} 
aspects also complicate the
creation of secure
client-side browser tools, extensions, and web-pages for the utilization of CT. 

To compare root stores, the tool removes duplicates from JSON root lists returned by {\it get-roots}, leaving only distinct certificates (see ``Distinct certificates" and ``Duplicated certificates" in Table~\ref{tab:logs2}). To identify equivalent root stores, we concatenate unique root hashes in ascending order and compare the obtained fingerprints (see ``Distinct list"). 
\revthree{}{Over time we have seen an increase in the number of logs (54$\rightarrow$57), but a decrease in the number of distinct lists (18$\rightarrow$15).}

\revtwo{To determine status and expiry constraints of logs we performed test submissions 
(see ``Test submission" in Table~\ref{tab:logs2}). For each log, a single trusted root certificate was submitted.}{To determine status and expiry constraints, for each log, we performed several test submissions of trusted (expired/non-expired) certificate chains.
\revthree{(see ``Test submission" in Table~\ref{tab:logs2}).}{}}
We then determined whether the submissions were 
successful (``$+$")
or unsuccessful (``$-$") due to an expiration rejection criteria 
\revthree{(``EXP") or}{or some other reason.} 
\revtwo{due to a failure associateda failure associated with an HTTP response code (``400" -- Bad Request, ``503" -- Service Temporarily Unavailable)} As expected, logs with a year in their name reject certificates that do not expire within a specified period. 
For the 
\revthree{rest of the}{other} 
logs, our 
test submissions were successful except for 
\revthree{Google Aviator and Venafi Gen2 CT.}{two logs (Aviator and Venafi Gen2 CT) in Dec. 2018 and eight logs in Oct. 2019 (Nimbus20\{17-18\}, Argon2017, Nessie/Yeti2018, Aviator, Venafi Gen 2 CT, WoTrus; five logs by CNNIC, GDCA, and SHECA are now offline or have an expired certificate).} 
\revthree{}{Some logs reject expired certificates (see ``Reject expired" in Table~\ref{tab:logs2}); logs with expiration constraints may be listed as ``usable" (e.g. Argon2018), but by rejecting expired certificates they essentially become frozen.}
\revthree{}{In some cases, DigiCert Log Server 2 rejects 
submissions due to
rate limitations.}



\section{High-level Comparisons}\label{sec:roots}

We next characterize high-level relationships between root stores.
For detailed, interactive exploration of these relationships, 
\revthree{we encourage the reader to use our supplied tool.}{we encourage the use of our tool.}

\begin{figure*}[t]
   \centering
   \includegraphics[trim = 0mm 18mm 0mm 20mm, width=0.82\textwidth]{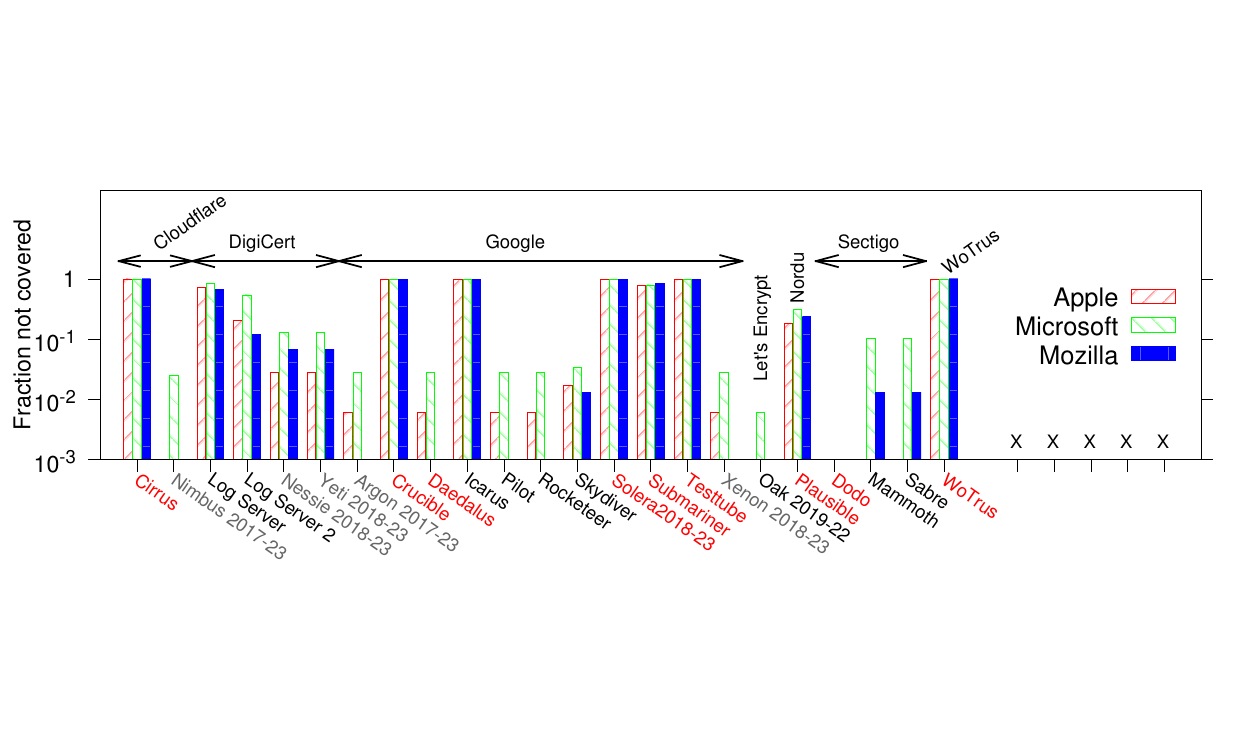}
   \includegraphics[trim = 0mm 18mm 0mm 20mm, width=0.82\textwidth]{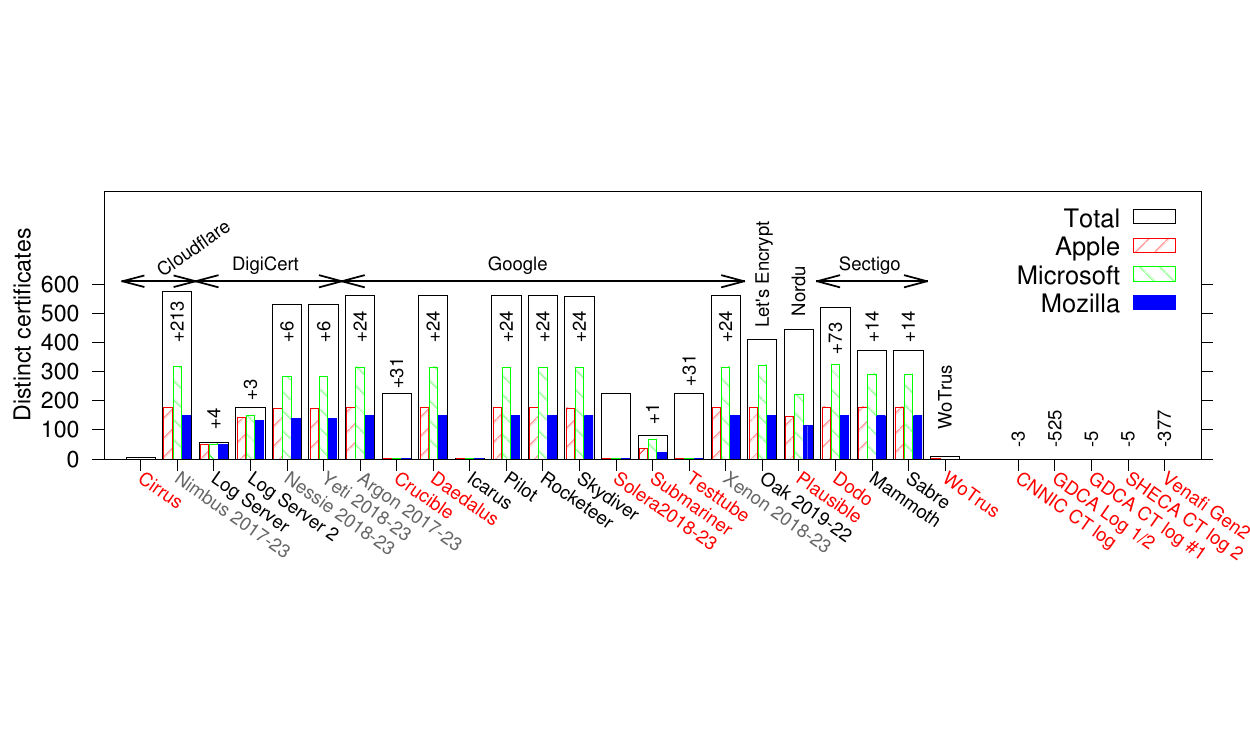}
   \vspace{-14pt}
    \caption{Top: Fractions of vendor root stores that are not covered by the logs.  Bottom: Log root stores and intersections with vendor root stores. Positive/negative numbers designate change between the initial and final measurement. \revthree{}{Log name color: ``Trusted according to Google's list of trusted logs" (black), ``Not Trusted/Not included in Google's list of trusted logs" (red), ``Trustworthiness depends on a year" (grey).}}
   \label{fig:store-coverage}
   \vspace{-6pt}
\end{figure*}

{\bf Vendor coverage:}
Figure~\ref{fig:store-coverage} (top part) shows the fraction of roots trusted by Mozilla, Microsoft and Apple that are not covered by the logs (using logarithmic scale).  A closer look reveals 
interesting observations.
First, the root stores are significantly different both between and within individual operators.
\revfour{}{As expected, for individual operators, the vendor coverage is typically (but not always) highest for trusted logs and smallest for testlogs and other non-trusted logs (e.g., Cirrus).}
\revthree{\revthree{In total, the 55 studied logs (excluding Aviator, which has no roots) use 18 distinct root stores (see ``Distinct list" in Table~\ref{tab:logs2}).}{In total, the 57 studied logs (excluding logs without roots) use 15 distinct root stores (see ``Distinct list" in Table~\ref{tab:logs2}).}}{}
\revthree{Second, despite the CT standard suggesting that a list of acceptable root certificates ``might usefully be the union of root certificates trusted by major browser vendors"~\cite{RFCCT},}{Second, motivated by the CT standard suggesting that a list of acceptable root certificates ``might usefully be the union of root certificates trusted by major browser vendors"~\cite{RFCCT}, 
many logs have increased their coverage to 100\% for at least one or two vendors.  (Changes are shown in Table~\ref{tab:logs2} and support for our ``motivated by" claims are based on our operator survey in Section~\ref{sec:survey}.)  Third, despite this increase,}
every log 
\revthree{}{except Dodo (72 added roots),}
is missing some fraction of roots trusted by 
\revthree{Mozilla, Microsoft and/or Apple.}{Mozilla, Apple or Microsoft.}  
\revthree{Third,}{Fourth,} 
on average, the root stores of Mozilla and Apple are better covered than Microsoft's. 
\revthree{Fourth, logs by Sectigo have the highest coverage for all three vendor stores. They contain 99.3\% and 100\% of roots trusted by Mozilla and Apple, respectively, and Sectigo Dodo covers 99.7\% of Microsoft's roots.}{Fifth, DigiCert, 
has the lowest vendor coverage of the five trusted log operators, and do not appear to have adjusted its root stores to match the changes made by the vendors.}

\begin{figure}[t]
  \centering
\includegraphics[trim = 0mm 2mm 7mm 4mm, width=0.57\textwidth]{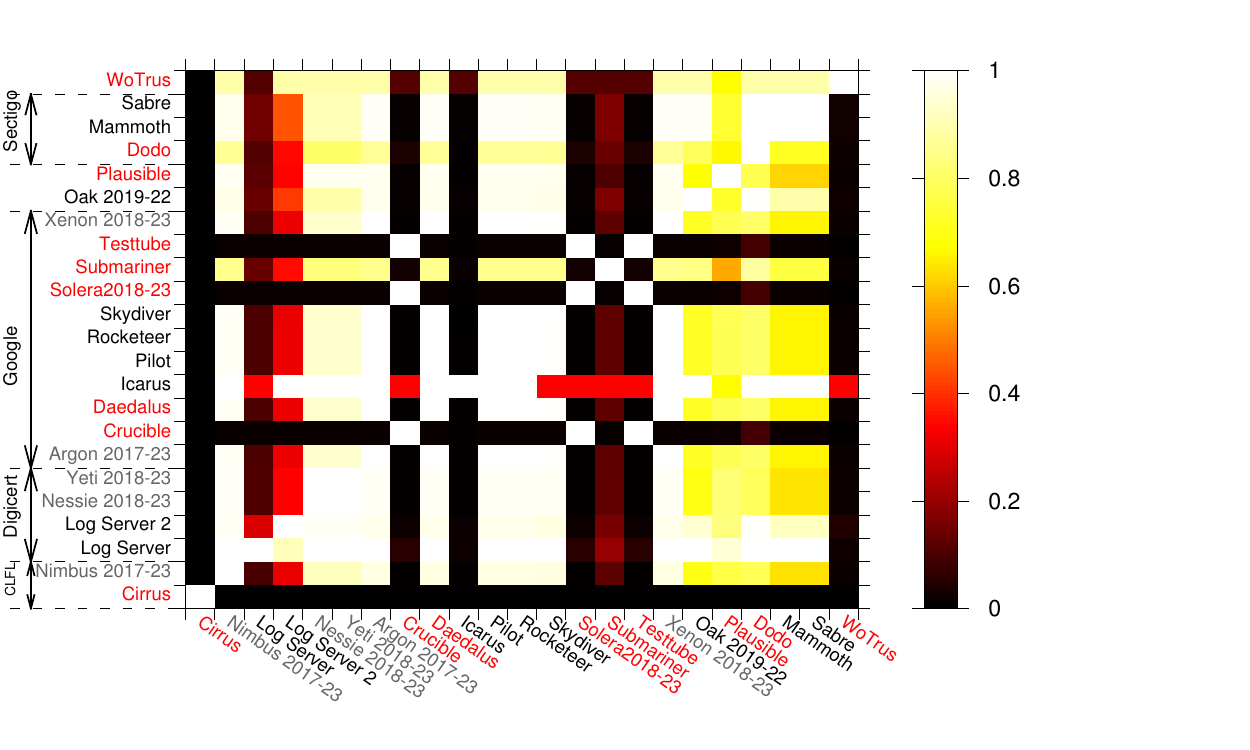}
  \vspace{-25pt}
  \caption{Pairwise root store overlap between CT logs. $\left | LogX\cap LogY \right | / \left | LogX \right |$. 
  \revthree{}{Text color label: ``Trusted" (black), ``non-trusted" (red), depends on expiry date (grey).}}
  \label{fig:root-store-overlap}
  \vspace{-6pt}
\end{figure}

{\bf Trusted vs. non-trusted logs:}
Although Apple's list of trusted logs differs from that of Google 
\revthree{(see ``Vendor trust" in Table~\ref{tab:logs2}),}{(see respective ``Log Program" in Table~\ref{tab:logs2}),} 
the differences are small and mostly related to logs with expiration dates in 2017-2018.
\revthree{}{We also illustrate this in Figures~\ref{fig:store-coverage} and~\ref{fig:root-store-overlap}, where we differentiate between ``trusted" logs (black text), ``non-trusted" logs (red text), and the production logs (e.g., Nimbus 20\{17-23\}) that are trusted for years 20\{19-23\} but no longer for the prior year (e.g., 20\{17-18\}).}
Most importantly, both Apple and Google rely exclusively on logs by 
\revthree{four operators: Cloudflare, DigiCert, Google and Sectigo.}{five operators: Cloudflare, DigiCert, Google, Sectigo, and Let's Encrypt (while the last operator's logs became Google ``qualified" on Oct. 7, 2019).} 
Logs from other operators  
\revthree{are either frozen (i.e., Venafi Gen2 CT) or}{are} 
\revthree{non-trusted.}{non-trusted (i.e., rejected, retired, not listed, or read-only).}  
\revthree{Furthermore, every Apple-trusted log, apart from the Sectigo Mammoth and Sabre, violates Apple's requirement to ``trust all root CA certificates included in Apple's trust store"~\cite{CTPolicyApple}.}{Furthermore, no Apple-trusted log by Google or DigiCert satisfies Apple's requirement to ``trust all root CA certificates included in Apple's trust store"~\cite{CTPolicyApple}, and both Cloudflare and Let's Encrypt only do so because of recent updates to their root lists.}
These observations suggest that some requirements of the evolving policies are not strictly followed.
\revthree{Among Chrome-trusted logs, Google Aviator and Icarus stand out,
as Aviator's root list does not contain any roots (frozen since Nov. 2016) and Icarus only has a few (bottom of Figure~\ref{fig:store-coverage}). 
Among}{Among} 
non-trusted logs, we note many test logs (e.g., Testtube, Solera). Their root lists are smaller, mainly contain test certificates, and miss almost all roots included by major vendors.

  \begin{figure}[t]
  \centering
  \includegraphics[trim = 5mm 0mm 2mm 5mm, clip, width=0.49
  \textwidth]{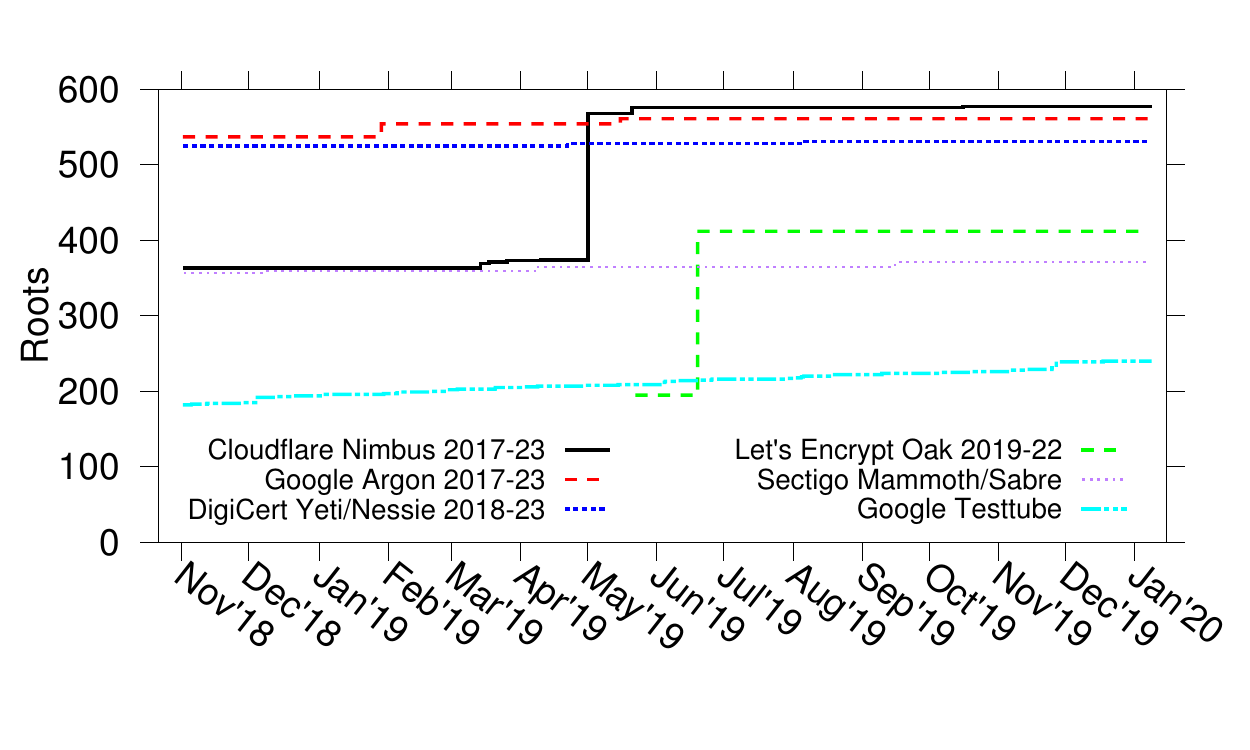}
  \vspace{-40pt}
  \caption{Root store sizes as a function of time.}
  \label{fig:time-sizes}
  \vspace{-6pt}
  \end{figure}
 
\revthree{}{{\bf Root store size evolution:} Figure~\ref{fig:time-sizes} shows how the root store sizes of the five major log operators have 
\revfour{changed over time.}{changed over time; for the most part making the root stores larger and more similar in size.} 
We also include the timeline of a testlog (Testube, which shares root store with Crucible and Solera 20\{18-22\}).  
Cloudflare (May 2019) and Let's Encrypt (June 2019) both made major increases to their root stores, making them more similar in size to those of Google and DigiCert.  At this time, they also increased their coverage of the root stores of the major vendors.
(From the survey, we learned that Cloudflare update in May 2019 was triggered by an email from a CA.)
\revfour{Today, only Sectigo Mammoth/Sabre have substantially smaller root store. (Interestingly, they still have the highest vendor coverage.)}{Interestingly, despite Sectigo Mammoth/Sabre having the highest vendor coverage (Figure~\ref{fig:store-coverage}), they have the smallest root store.} 
\revfour{We also}{Finally, we} 
note that during our measurement campaign the production logs had much fewer root store change events (three on average) 
than the testlog (41 events).}
 
 \begin{figure*}[t]
  \centering
  \subfigure[Dec. 27, 2018]{
    \includegraphics[trim = 32mm 55mm 30mm 50mm, width=0.46\textwidth]{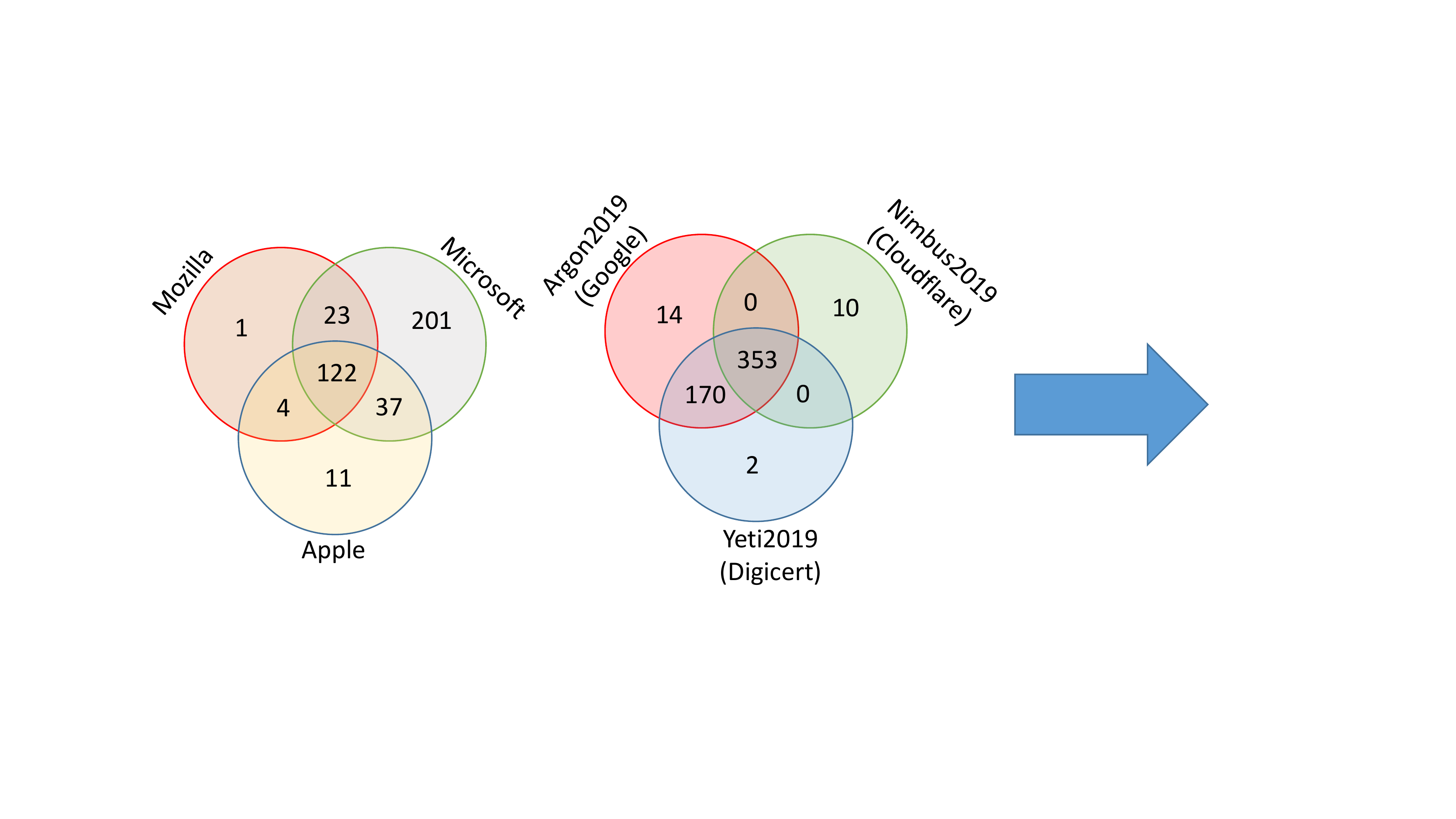}}
    \subfigure[Oct. 8, 2019]{
    \includegraphics[trim = 32mm 55mm 38mm 50mm, width=0.48\textwidth]{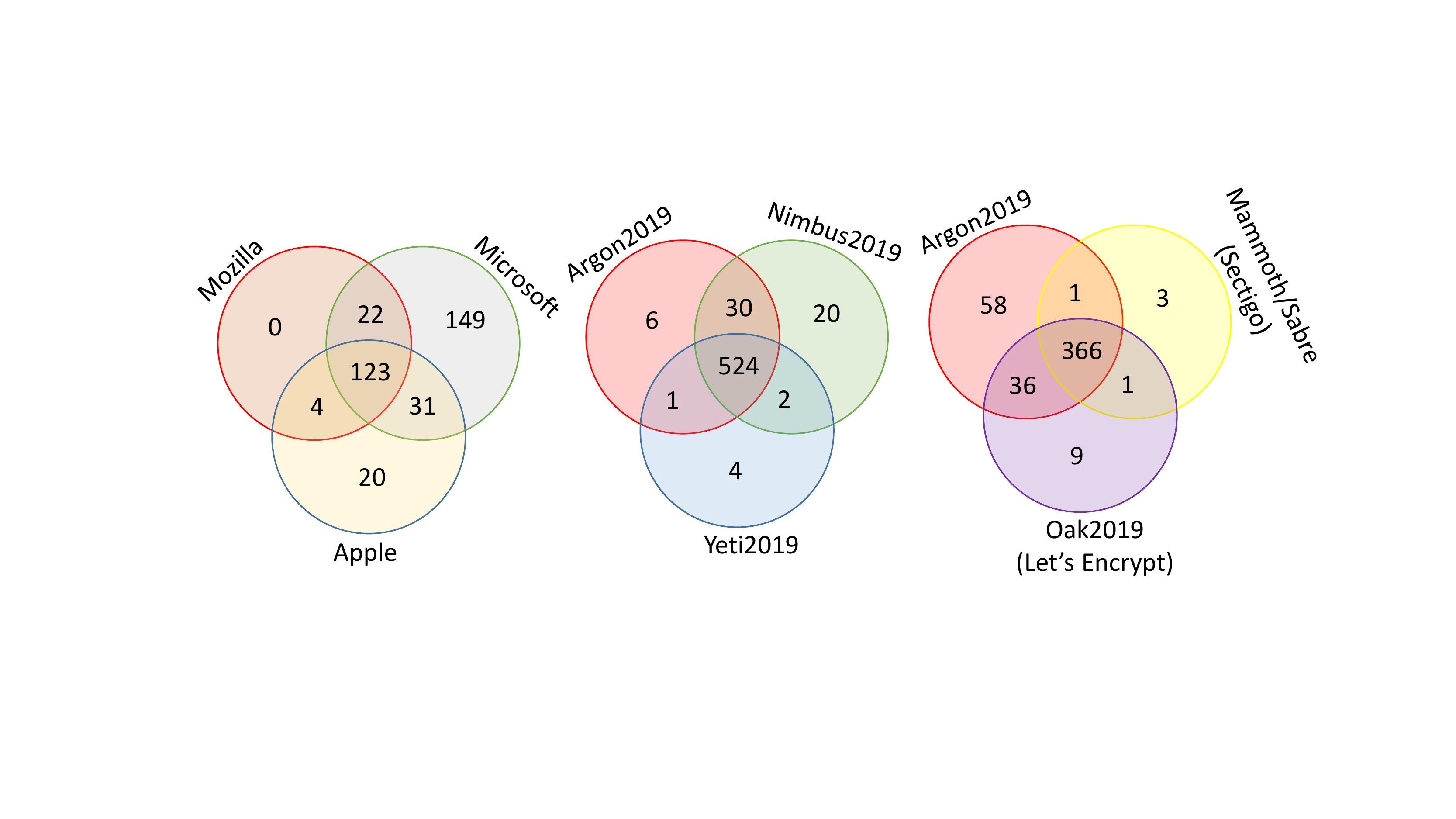}}
    \vspace{-6pt}
    \caption{Intersections of vendor root stores and example production logs.}
    \label{fig:venn}
      \vspace{-8pt}
\end{figure*}
 
{\bf Root store (dis)similarity:}
Figure~\ref{fig:root-store-overlap} shows the pairwise root store overlap between logs, where the overlap is calculated as $\left | X\cap Y \right | / \left | X \right |$ for each pair of logs $X$ and $Y$. 
We note that many pairs overlap substantially and differ only by a few roots (light yellow).
The largest differences are between the test logs and production logs (e.g., darker regions associated with Google Solera, Crucible, and Testtube).
Previously, two GDCA logs 
\revthree{}{(no longer accessible)}
and ten DigiCert logs
\revthree{are only}{were} 
``identical" after removing duplicates.
\revthree{Finally, we highlight some differences observed between the vendor stores (top of Figure~\ref{fig:venn}) and the three major Chrome/Apple-trusted logs (bottom of Figure~\ref{fig:venn}). In each case, the individual root stores contain unique roots.}{Finally, Figure~\ref{fig:venn} highlights some differences observed between the vendor stores and the major Chrome/Apple-trusted logs, and how the overlaps between some of these root stores have changed.  In almost all cases, the individual root stores contain unique roots.} 
Among the vendors, Microsoft has the largest and most dis-similar root store.  
\revthree{Among the logs, Google Argon and DigiCert Yeti/Nessie logs share most of the roots with each other, whereas Cloudflare Nimbus 
trusts 
two-thirds of that.}{Among the logs, the root store of Cloudflare Nimbus became much more similar to the root stores of Google Argon and DigiCert Yeti/Nessie, who already shared most of their roots.  
Let's Encrypt Oak and Sectigo Mammoth/Sabre have fewer roots and smaller overlap with Argon.  As a reference point, we remind the reader that Mammoth/Sabre has close to full coverage of the three vendor stores.}

\begin{figure*}[t]
   \subfigure[Dec. 27, 2018]{\includegraphics[trim = 13mm 21mm 27mm 20mm, clip, width=0.48\textwidth]{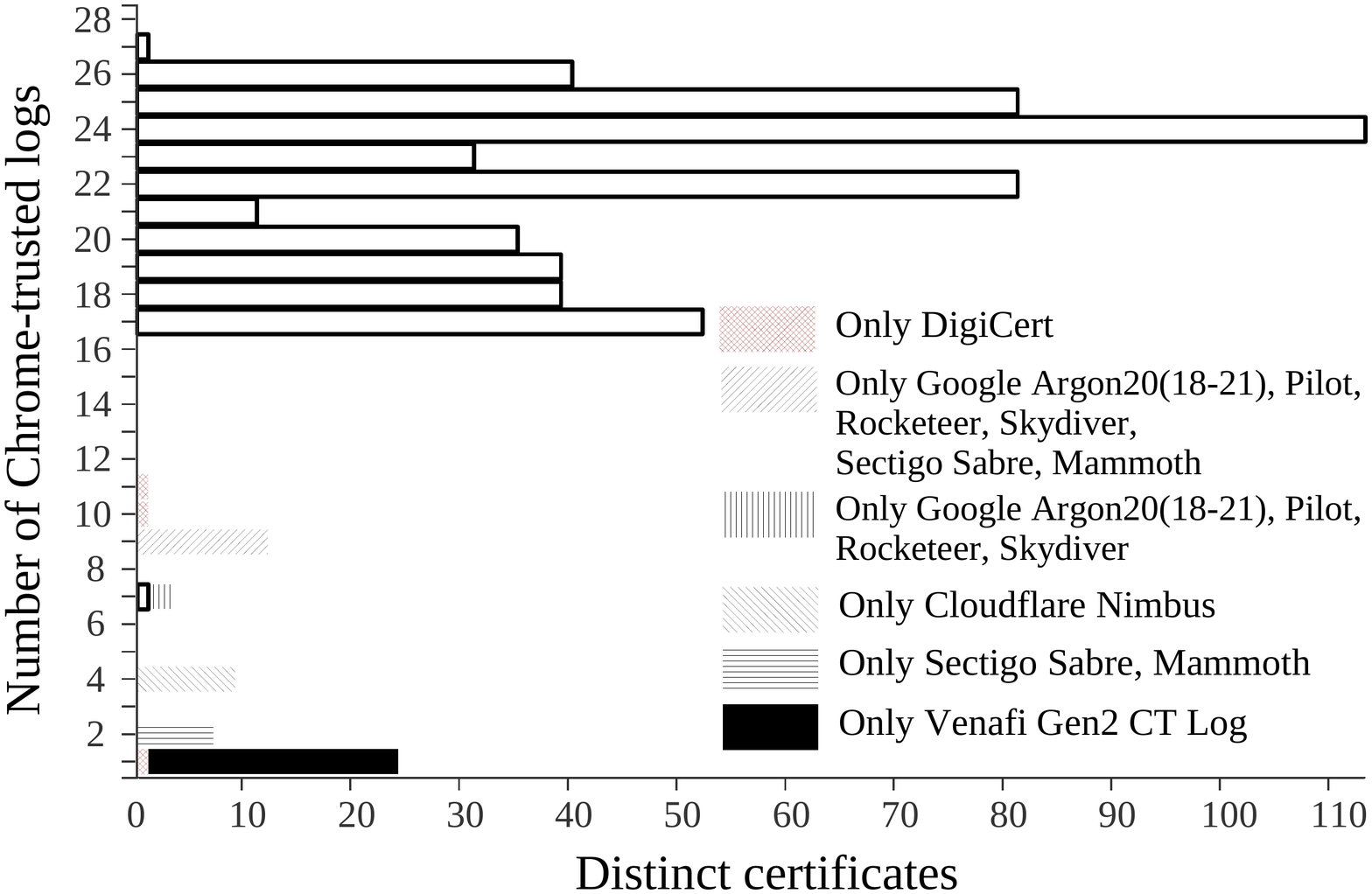}}
\subfigure[Oct. 8, 2019]{\includegraphics[trim = 13mm 21mm 27mm 20mm, clip, width=0.48\textwidth]{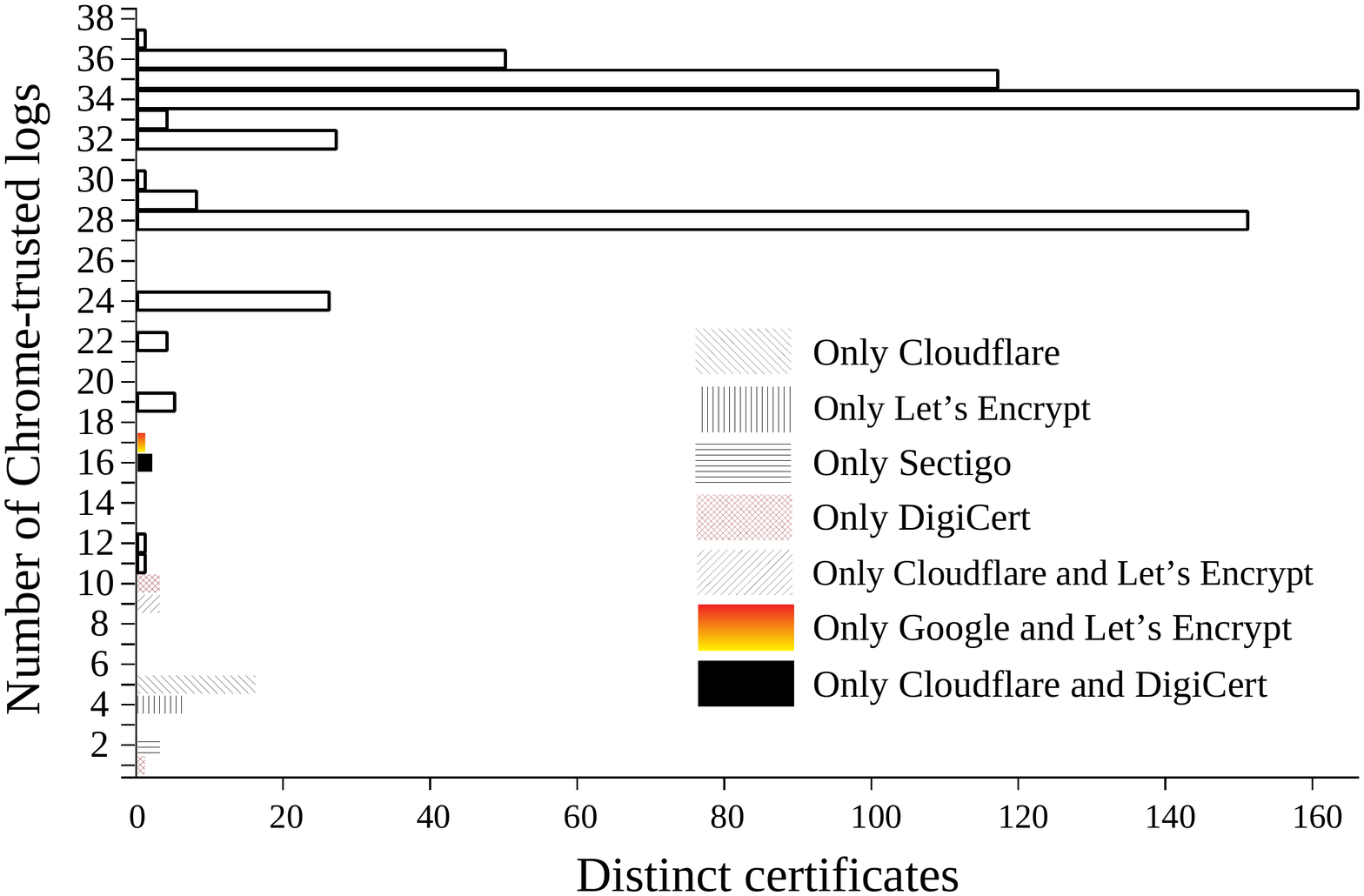}}
   \vspace{-6pt}
   \caption{Certificates grouped by how many Chrome-trusted logs include 
   \revthree{them in their root lists}{them.}}
   \label{fig:frequency}
      \vspace{-6pt}
\end{figure*}

{\bf Root frequencies:}
The trustworthiness of CAs is relative.  It is therefore 
\revfour{natural}{not surprising} 
that some roots are more frequently included than others. Figure~\ref{fig:frequency} shows the frequency distribution of the observed 
\revthree{roots.}{roots seen at the time of our two main snapshots.} Here, we first grouped distinct certificates by the number of Chrome-trusted logs that use them as roots, and then plotted the number of certificates in each group.
The most frequent root certificate 
\revfour{\revthree{(trusted by all 27 Chrome-trusted logs, excluding Aviator, which has an empty root list)}{(trusted previously by 27 and currently by 37 Chrome-trusted logs)}}{(previously included in the root lists of 27 Chrome-trusted logs and currently by 37 such logs)}
is the {\it Merge Delay Monitor Root}. 
Logs are required to include this root to become Chrome-trusted, but the root is also present in most non-trusted logs. 
\revthree{When Apple introduce its own monitoring root CA yet, it is likely going to become one of the most frequently included roots too.}{Apple's corresponding root is not as frequent.}
Otherwise, the certificates fall into two categories. Most of the vendor roots are covered by 
\revthree{17-26}{the majority of the} 
Chrome-trusted 
\revthree{logs and}{logs.  Except for the two roots (covered by 11 and 12 logs in the last snapshot),}
the rest of the roots covered by less than 17 logs (some of which are not present in vendor stores) are 
\revfour{trusted}{included in the root lists} 
by the logs of just one or two operators. 
\revthree{Except for 
\revthree{one root ({\it UTN-USERFirst-Object}),}{two roots ({\it UTN-USERFirst-Object} and ????),} 
all less frequent root certificates are trusted by just one or two log operators, which is not sufficiently redundant. The}{This is not sufficiently redundant.} 
\revthree{The list of affected issuers includes Athens Stock Exchange, Camerfirma, certSIGN, DigiCert, 
DigiNotar (discussed in Section~\ref{sec:root-standouts}), emSign, Entrust, Google, GlobalSign, Globaltrust, Halcom, LGPKI, Microsoft, Netrust, Skaitmeninio sertifikavimo centras, Thawte, T-Systems, and WISeKey (the list is not exhaustive). {\it UTN-USERFirst-Object} is included in Microsoft and Apple root stores (not Mozilla) and trusted by seven logs from three operators (Cloudflare, DigiCert and Sectigo). 
Venafi Gen2 CT (now frozen) is responsible for the majority of the least frequent certificates; however, most of these certificates are not self-signed roots, but intermediaries that can be chained back to already included roots.}{Venafi Gen2 CT, responsible for the majority of the least frequent certificates in the Dec. 2018 dataset, is now frozen. However, most of these certificates are not self-signed roots, but intermediaries that can be chained back to already included roots.}

  \begin{figure*}[t]
  \centering
  \subfigure[Roots used each day]{
  \includegraphics[trim = 2mm 10mm 2mm 7mm, width=0.48\textwidth]{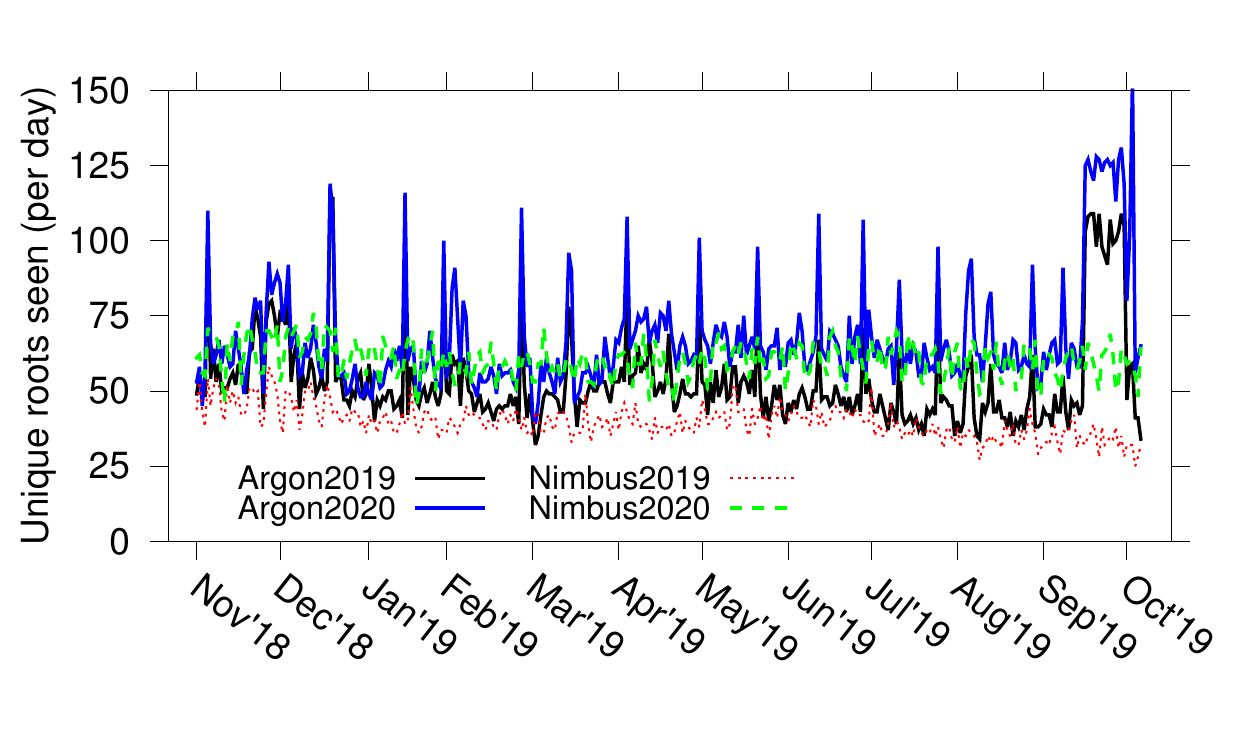}}
  \subfigure[Days root used]{\includegraphics[trim = 0mm 10mm 0mm 2mm,  width=0.48\textwidth]{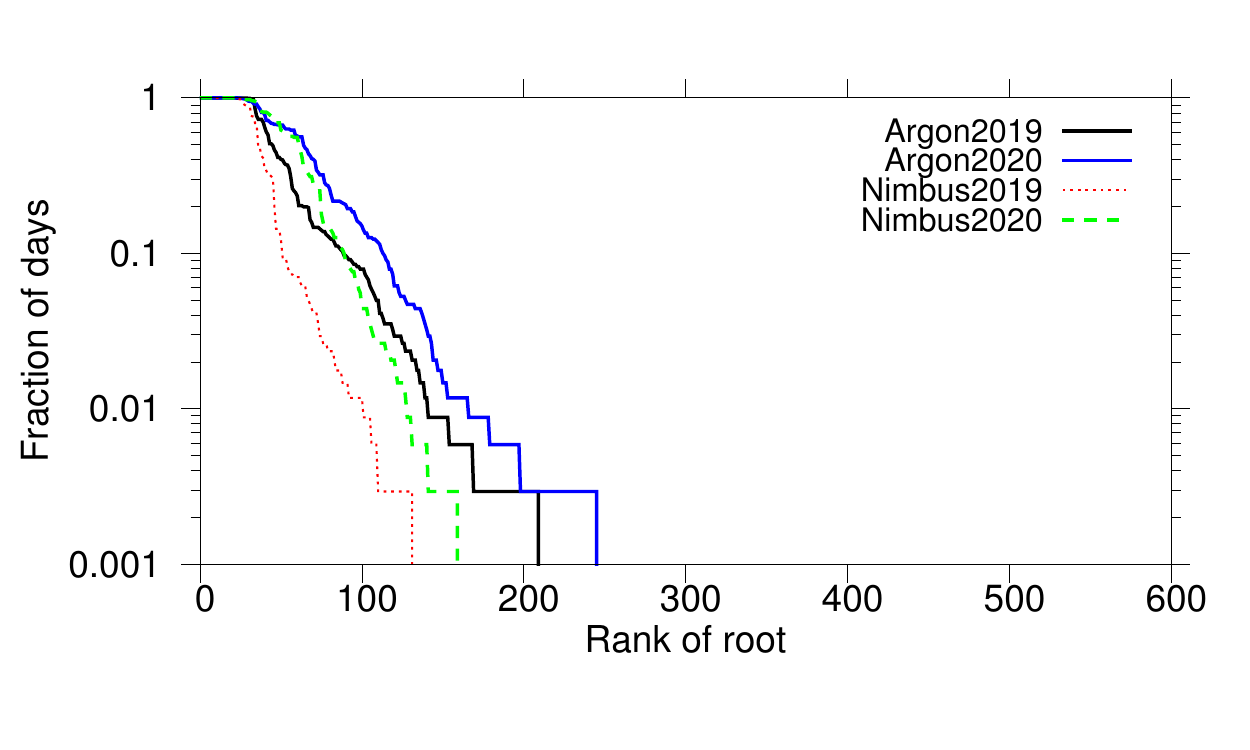}}
  \vspace{-6pt}
  \caption{Upper bounds on individual root usage.}
  \label{fig:root-usage}
  \vspace{-6pt}
  \end{figure*}

\revthree{}{{\bf High skew in root usage:}  There is a high skew in the usage of different roots and many of the roots included by the major logs are not used.  Figure~\ref{fig:root-usage}(a) shows an upper bound on the number of roots that was used by the Nimbus20\{19,20\} and Argon20\{19,20\} logs each day during the measurement period and Figures~\ref{fig:root-usage}(b) shows an upper bound on the number of days that each root may have been used.  These bounds were obtained using Censys~\cite{censys15} by extracting information about all certificates logged to these logs and checking which set of roots that these certificates chain back too (can be more than one).  Comparing these numbers with the root store sizes of these logs (which changes from 363 to 576 and from 537 to 561, respectively, during the measurement period), we note that more than half of the roots are not used at all, and less than 113 (Argon2020) of the roots are used more than 10\% of the days.}


\section{Log management}\label{sec:management}

\subsection{Operator survey}\label{sec:survey}

\revthree{}{We emailed questions (Sept. 23, 2019) about log management to the five Google/Apple trusted log operators, and got answers from all five: Google, Let's Encrypt, Cloudflare, Sectigo, and DigiCert.  The questions are listed in appendix.}

\revthree{}{With exception of DigiCert, the operators create their root lists automatically, in some cases involving a manual review step before lists are 
promoted to the production logs. Both Google and Sectigo explicitly state that they use the union of the trusted roots from the Mozilla, Microsoft, and Apple trusted root programs.  The main difference is that Google typically do not remove roots from the set of accepted roots (although they state that there is on-going discussion about potentially starting to do this), whereas Sectigo base their lists on their current crt.sh data~\cite{SectigoRoots2019}.  Cloudflare have their own roots program, in which they explicitly lists root stores they cover, but they also 
update based on requests from CAs.  Let’s Encrypt also mentioned that they obtained a root from Apple (per email) to perform similar tests as Google’s MMD root and that they plan to create similar tools as those used by Cloudflare (e.g.,~\cite{CloudflareRoots2019}).} 

\revthree{}{DigiCert maintains a master file of 
\revfour{trusted}{acceptable} 
roots that gets pushed out with every deploy. The original list for their high performance logs was created from the browser trust stores, to which they add roots included in at least one browser root store as email requests to include these come in.  At some point, they plan to find and add all the roots that are acceptable by the new Google logs.}

\revthree{}{Cloudflare and Sectigo stated that (as far as they know) they have not intentionally enabled or disabled CORS, with Cloudflare noting that they currently do not set them since they have not yet had reason to update their code (which is based on the Trillian code base) to the latest version of Trillian (which allows CORS headers), and note that such updates would involve significant changes and testing of closed-source code for their internal infrastructure.} 

\revthree{}{All operators seem to agree that the process of root management is secure enough.  The main comments that came up were related to more transparent root management of the logs. For example, Let’s Encrypt mentioned the value of public monitors such as the one developed (and made available) in this paper and how they believe that the CT landscape could benefit from root program operators providing an easily ingestible list of root CAs trusted on various platforms.  DigiCert suggested that it would be nice if log operators had a statement explaining their criteria for including new roots.  For future versions of CT, Google suggests to
\revfour{requiring logs to commit changes to their set of trusted roots as a new leaf type, which they log to themselves.}{``require logs to commit changes to their set of trusted roots as a new leaf type which they log to themselves."}  This would make it harder to exploit logs via root sets, and would provide a new mechanism for others to become aware of updated root sets.}

\subsection{Incidents and potential mismanagement}\label{sec:mismanagement}

\revthree{Over the period of our study we}{We} 
\revfour{have observed a number of events and behaviors that warrant comments.}{made a number of additional observations.}

{\bf Duplicated roots:} 
Several logs respond to the {\it get-roots} method with 
root lists that include duplicates 
\revthree{(Table~\ref{tab:duplicates}).}{(``Duplicates" in Table~\ref{tab:logs2} and Table~\ref{tab:duplicates} in appendix).} 
These duplicates may have been introduced by accident due to manual log (mis)configuration, 
or a bug in a log server implementation. However, we have found that the misbehavior of GDCA and Sectigo logs (described next) coincided with the anomalous presentation of their root 
\revfour{stores.}{stores, when duplicate roots were introduced.} 
Moreover, the frozen Venafi Gen2 CT log also contains 
duplicates.
We conclude that the presence of duplicated roots could be used as an
indicator of log's 
\revfour{trust mismanagement.}{potential (root) mismanagement.} 
At this moment, DigiCert Log Server 2, Yeti and Nessie logs still contain duplicates.
\revthree{We suggest}{DigiCert acknowledge this as an oversight likely caused by a CA having asked for inclusion of a root that already was in the list of acceptable roots, but do not consider this a serious issue. We believe}
that a mechanism for transparent management of 
\revfour{trusted certificates}{root lists} 
\revthree{is required.}{would be beneficial.} 
Such a mechanism would improve monitoring and could prevent trust-related log misbehavior. Versioning of root stores in public repositories similar to~\cite{SectigoRoots} could be a possible solution.
 
{\bf GDCA logs:}
On Aug. 16, 2018, GDCA Log 1 failed to incorporate two SCTs within the MMD. The SCTs were issued during an update of the root lists for Log 1 and Log 2~\cite{GDCAaccident}.  According to 
GDCA,
the accident 
was caused by a reboot after the update.
Google disqualified Log 1 and GDCA withdrew their request to include Log 2 in Chrome. 
Interestingly, we found that
duplicates were introduced at the time of the incident and that they remained there until the logs were shut down (on Jan.~15 and Feb.~14, 2019).
\revthree{Furthermore, clients using Mozilla or Apple root stores cannot establish secure TLS connections to these logs, as they either use expired certificate (CT Log \#1) or do not chain to any of the roots in the vendor root stores (Log 1 and Log 2).}{}

{\bf Sectigo Mammoth, Sabre and Dodo:}
We have observed outages for Sabre and Mammoth, with up-time of Sabre going below 99\%. We have also noticed that Mammoth was sporadically returning two different root lists, as {\it Fina Root CA} and {\it Hongkong Post Root CA 3} have been randomly appearing in Mammoth's root list between Dec. 6, 2018 and Jan. 2, 2019. \commentout{These certificates are  
\revthree{not currently present in the root stores of these logs,}{currently not in the logs' root stores,} 
but are 
\revthree{both trusted}{trusted} 
by Dodo and Microsoft.}
\revthree{}{While Sectigo answered our survey, they have yet to comment on these events.}




%


\subsection{Notable roots and use cases}\label{sec:root-standouts}

{\bf Test roots in trusted logs:}
In general, you would not expect test roots in Chrome or Apple-trusted logs. However, we have found that DigiCert Log Server 2, Yeti and Nessie logs 
\revfour{trust two {\it DigiCert CT Test Root}s that are not directly logged in any of the trusted logs.}{include two {\it DigiCert CT Test Root}s in their root lists that are not directly logged in any of the trusted logs.} 
\revthree{(We have resisted the temptation to log the certificates ourselves!)}{(While most attempts were rejected due to rate limits setup for the older DigiCert logs, to protect the log from risk of failure or missing the MMD for new entries, we could not resist the temptation, and have now logged one of these certificates~\cite{logged-cert-2019}.)} 
\revthree{}{DigiCert have since explained that they included some test roots for monitoring purpose.}
Let's Encrypt logs also include a test root – {\it ct-woodpecker}.

{\bf DigiNotar Root CA in Cloudflare logs:}
DigiNotar is a former CA 
that went bankrupt due to a well-known attack~\cite{prins2011diginotar} on its infrastructure.
This attack and the fact that misissued certificates could remain undetected was one of the reasons for the creation of CT~\cite{AfterDigiNotar}. 
Surprisingly, Cloudflare Nimbus logs include {\it DigiNotar Root CA} certificate in their lists of acceptable roots, years after the attack. 
\revfour{According to the public incident report~\cite{prins2011diginotar} this root was compromised, but no evidence suggesting its misuse was found. However, assuming}{Assuming} 
that there still exists a party with access to 
its private key,
this party could issue an arbitrary number of DigiNotar-signed certificates acceptable by Nimbus logs\footnote{\revfour{A successful attack on a log may interrupt issuance process of some CAs.}{While there exist many other ways to obtain numerous valid certificates for submissions (e.g., downloading certificates from other logs or tweaking existing ones), the capability to generate arbitrary many acceptable certificates could also be used for DoS attacks.
We note that a successful attack on a log may interrupt issuance process of some CAs relying on that log.}}.
\revthree{}{When asked, Cloudflare did not know why this root was in their root store, but have no plans to remove it either.  
\revfour{}{(A closer look at the Cloudflare root policy~\cite{CloudflareRoots2019} suggests that the inclusion may be due to them using the union of several OS trust stores, including some that include DigiNotar (e.g., Android Gingerbread and Honeycomb).}
While we have seen roots being removed, some operators appear to rather be inclusive than hinder some CAs from submitting.}

\revfour{}{The above example presents an interesting dilemma. Nowadays, Cloudflare Nimbus is the only log that allows DigiNotar certificates. It potentially could help understand or even identify the attacker if a rogue certificate is logged. However, in the case when other operators add this certificate to their lists of acceptable roots (e.g., to increase the chance that such certificates are logged), an attacker would be able to obtain enough SCTs to satisfy requirements of CT policies, and hence {\it may} more easily trick a client on which the DigiNotar Root CA certificate is somehow valid. To avoid such future conflicts, we suggest that non-trusted roots perhaps should be stored in separate logs; e.g., similar to how Google Daedalus (discussed below) is used to store expired certificates. Moreover, a common and homogeneous approach to the formation of root stores in logs would eliminate such outlying cases.}


{\bf Cirrus:}
\revfour{Rather}{A closer look at the root stores (and the logs' content) also highlight other interesting example use cases of CT.  For example, rather}
than storing certificates associated with the WebPKI, Cloudflare Cirrus is used with Resource Public Key Infrastructure (RPKI)~\cite{CirrusBlog} and its root store only contains the certificates of the five Regional Internet 
\revfour{\revthree{Registries (RIRs): AFRINIC, APNIC, ARIN, LACNIC, RIPE.}{Registries.}}{Registries (RIRs): AFRINIC, APNIC, ARIN, LACNIC, RIPE.}

{\bf Daedalus:}
While the root store of Google Daedalus is equivalent to the stores of the Argon, Xenon, Pilot and Rocketeer logs, it is unique in that it only accepts expired certificates.  It is not trusted by Chrome nor Apple.
\revfour{}{We believe that Daedalus provides an example of how non-trusted logs can be used to log non-trusted certificates of potential interest, which we believe could be extended to cover some of the roots that the browsers no longer trust (e.g., DigiNotar).}

{\bf SHECA CT log 2:}
\revfour{This log}{The skew in usage of the logs is high~\cite{FirstLookCT}, with some logs seeing very limited use. As an extreme example, SHECA CT log 2}
was idling with just two certificates in its tree between Apr. 2017 and Jan. 2019, when we 
successfully performed a test submission.
The log's server certificate 
\revthree{can}{could} 
be chained back to Mozilla's root store, but not to 
\revthree{Apple's trust 
store,}{Apple's,} 
meaning that by default,
Apple clients 
\revthree{are}{were} unable to establish a secure connection to the log.
\revthree{}{The log was taken offline in Mar. 2019.}

\section{Related work}\label{sec:related}

While a number of studies have characterized the CT 
landscape~\cite{FirstLookCT,RiseOfCT},
very limited results have been reported regarding their root selections.
For example, Gustafsson et al.~\cite{FirstLookCT} provide basic root count for 
\revthree{11}{eleven} 
public logs available in Dec. 2015 and list basic log properties, but primarily provide statistics (based on active and passive measurement data) of the relative CT usage among domains and CAs.
Most other papers focus on logged certificates~\cite{InLogWeTrust,kumar2018tracking} and on specific aspects, such as server-side use of SCTs~\cite{NSGC18,AfterDigiNotar} and client-side performance of SCT delivery methods~\cite{NSGC18}. 
Amann et al. have studied and compared the adoption of several technologies (including CT) that strengthen the WebPKI~\cite{AfterDigiNotar}.
Privacy issues of CT are considered by Eskandarian et al.~\cite{eskandarian2017certificate}. A longitudinal study on CAs, log operators, and CT deployment on servers is presented by Scheitle et al.~\cite{RiseOfCT}. In the study, they also apply CT for the detection of phishing domains and highlight the use of CT by third-parties for malicious purposes. 
Stark et al.~\cite{CTFreshAdoption} study the adoption of CT and estimate the compliance to Google's CT policy using client-side Chrome telemetry.
In contrast to prior work, our focus is on the root stores of the logs and their relationship to the root stores of major vendors and CT policies.



%
%
%
%
%
%
%

\section{Conclusion}\label{sec:conclusions}

In this paper, we presented the first characterization of the emerging root store landscape and our an interactive analysis tool -- {\em CT Root Explorer}.
We overviewed CT logs and analyzed their root stores relative to the stores of Apple, Microsoft and Mozilla. 
\revthree{We have summarized properties of 56 CT logs, including 54 available logs from Google's list of known logs.}{We have monitored and summarized the properties of 57 CT logs, including all available logs from Google's and Apple's lists of known logs, and observed how the root stores change over time.} The landscape of CT is evolving: the technology is highly established and is commonly used by the {\it 
majority} of CAs and clients on the Internet, which makes it strategically important; software vendors introduce and change their CT policies, new logs are established regularly, while the recent introduction of Cloudflare Cirrus extended the use of CT to ResourcePKI. 
\revthree{}{Over our measurement campaign root stores have became larger and increased their coverage of the roots used by the major vendors.} To directly measure the status of the available CT logs, we have performed a number of test submissions, and by calculating frequencies of root certificates in logs, we have found a number of roots that are 
\revfour{trusted}{included in the root lists} 
by just one or two log operators. Surprisingly, Cloudflare Nimbus logs 
\revfour{announce trust in a compromised DigiNotar root certificate.}{include a compromised DigiNotar root certificate in their root lists.} 
Multiple logs are considered Apple/Google-trusted despite violating corresponding policies. We have discovered that all CT logs (except Google's and Let's Encrypt's) do not specify cross-origin headers in their HTTP responses, which obstructs access to the logs using JavaScript in modern browsers. Overall, we have observed that some WebPKI roots trusted by major software vendors are not sufficiently covered by the CT logs; Apple and Google rely on the CT backbone that is comprised of just 
\revthree{4}{five} 
actively logging operators: 
Cloudflare, DigiCert, 
Google, and Sectigo (all of these operators are US-based).
\revfour{}{ Several logs have already been disqualified, while Sectigo's trusted production logs (Mammoth/Sabre) have been seen to struggle with their up-time requirements.}
\revthree{Most importantly, we}{Moreover, we} 
have found logs with duplicates in their root 
\revthree{lists. Moreover,}{lists and that} an accident with GDCA Log~1 and outages of Sectigo Mammoth coincide with anomalous presentation of their root stores. One other misbehaving log which is currently in a read-only state (Venafi CT Gen2) has also been presenting root stores with duplicated entries in it.
\revthree{Finally, we suggest}{Finally, we argue} 
that management of CT policies and logs' root stores must be performed in a more careful, 
timely, and transparent manner.










\section*{Acknowledgment}
We are very thankful to the log operators who responded to the survey and provided additional insights into our log specific observations.
This work was 
supported by the Wallenberg~AI, Autonomous Systems and Software Program (WASP) funded by the Knut and Alice Wallenberg Foundation.

\bibliographystyle{splncs04}
\bibliography{references}

\revone{
\begin{table*}[t]
  \centering
  \caption{Pairiwse comparsion of the number of non-overlapping certificates in the ``trust'' sets.}
  \label{tab:testFF}
  \vspace{-6pt}
  {\footnotesize
    \begin{tabular}{|c|c|c|c|c|c|c|c|c|c|c|c|c|c|c|c|c|c|}
      \hline
0 & \multicolumn{17}{l|}{GG1: Argon20XX, Xenon20XX, Pilot, Rocketeer, Daedalus (Google $\times$14)} \\\hline
537 & 0 & \multicolumn{16}{l|}{GG2: Aviator (Google)} \\\hline
534 & 3 & 0 & \multicolumn{15}{l|}{GG3: Icarusc (Google)} \\\hline
2 & 535 & 536 & 0 & \multicolumn{14}{l|}{GG4: Skydiver (Google)} \\\hline
194 & 363 & 362 & 194 & 0 & \multicolumn{13}{l|}{CF1: Nimbus20XX (Cloudflare $\times$7)} \\\hline
485 & 52 & 53 & 483 & 311 & 0 & \multicolumn{12}{l|}{DC1: Log Server (DigiCert)} \\\hline
367 & 176 & 173 & 369 & 235 & 134 & 0 & \multicolumn{11}{l|}{DC2: Log Server 2 (DigiCert)} \\\hline
16 & 525 & 522 & 18 & 182 & 473 & 353 & 0 & \multicolumn{10}{l|}{DC3+GD1: Yeti20XX, Nessie20XX, GDCA Log 1, GDCA Log 2 (DigiCert $\times$10, GDCA$\times$2)} \\\hline
206 & 377 & 376 & 206 & 156 & 337 & 305 & 194 & 0 & \multicolumn{9}{l|}{VE1: Gen2 CT log (Venafi)} \\\hline
196 & 357 & 354 & 198 & 202 & 305 & 203 & 208 & 280 & 0 & \multicolumn{8}{l|}{SE1: Sabre, Mammoth (Sentigo $\times$2)} \\\hline
479 & 80 & 81 & 477 & 355 & 116 & 200 & 471 & 399 & 313 & 0 & \multicolumn{7}{l|}{GG5: Submariner (Google)} \\\hline
725 & 194 & 195 & 723 & 551 & 240 & 364 & 713 & 547 & 545 & 270 & 0 & \multicolumn{6}{l|}{GG6: Solera20XX, Testtube, Crucible (Google $\times$7)} \\\hline
535 & 2 & 3 & 533 & 363 & 52 & 176 & 523 & 377 & 355 & 78 & 194 & 0 & \multicolumn{5}{l|}{GD2: CT log \#1 (GDCA)} \\\hline
186 & 449 & 446 & 188 & 228 & 397 & 277 & 202 & 302 & 92 & 391 & 605 & 447 & 0 & \multicolumn{4}{l|}{SE2: Dodo (Sectigo)} \\\hline
111 & 444 & 443 & 111 & 143 & 392 & 322 & 99 & 135 & 255 & 434 & 630 & 444 & 253 & 0 & \multicolumn{3}{l|}{NO1: Plausible (Nordu)} \\\hline
534 & 5 & 6 & 532 & 366 & 55 & 179 & 522 & 380 & 354 & 81 & 195 & 3 & 446 & 447 & 0 & \multicolumn{2}{l|}{SH1: CT log 2 (SHECA)} \\\hline
534 & 3 & 4 & 532 & 360 & 53 & 173 & 522 & 374 & 354 & 81 & 195 & 3 & 446 & 441 & 6 & 0 & \multicolumn{1}{l|}{CN1: CT log (CNNIC)} \\\hline
{\begin{sideways}{GG1}\end{sideways}} & {\begin{sideways}{GG2}\end{sideways}} & {\begin{sideways}{GG3}\end{sideways}} & {\begin{sideways}{GG4}\end{sideways}} & {\begin{sideways}{CF1}\end{sideways}} & {\begin{sideways}{DC1}\end{sideways}} & {\begin{sideways}{DC2}\end{sideways}} & {\begin{sideways}{DC3+GD1}\end{sideways}} & {\begin{sideways}{VE1}\end{sideways}} & {\begin{sideways}{SE1}\end{sideways}} & {\begin{sideways}{GG5}\end{sideways}} & {\begin{sideways}{GG6}\end{sideways}} & {\begin{sideways}{GD2}\end{sideways}} & {\begin{sideways}{SE2}\end{sideways}} & {\begin{sideways}{NO1}\end{sideways}} & {\begin{sideways}{SH1}\end{sideways}} & {\begin{sideways}{CN1}\end{sideways}} & \\\hline
   \end{tabular}}
   \vspace{-6pt}
   \end{table*}}{}

\revone{
\begin{table}[t]
\caption{Duplicated certificates}
\begin{center}
\begin{tabular}{|l|l|}
\hline
\textbf{Logs}                                                                                                 & \textbf{Duplicated Certificates}                                                                                                                 \\ \hline
\begin{tabular}[c]{@{}l@{}}DigiCert Log Server 2\\ DigiCert Yeti20* Log\\ DigiCert Nessie20* Log\end{tabular} & \begin{tabular}[c]{@{}l@{}}Atos TrustedRoot 2011\\ Atos, Germany\end{tabular}                                                         \\ \hline
\multirow{2}{*}{\begin{tabular}[c]{@{}l@{}}GDCA Log 1\\ GDCA Log 2\end{tabular}}                              & \begin{tabular}[c]{@{}l@{}}DigiCert Trusted Root G4 \\ DigiCert Inc, United States\end{tabular}                                       \\ \cline{2-2} 
                                                                                                              & \begin{tabular}[c]{@{}l@{}}Merge Delay Monitor Root\\ Google UK Ltd., \\ London, United Kingdom\end{tabular}                             \\ \hline
\multirow{6}{*}{Venafi Gen2 CT log}                                                                           & \begin{tabular}[c]{@{}l@{}}Buypass Class 2 Root CA\\ Buypass AS-983163327, Norway\end{tabular}                                        \\ \cline{2-2} 
                                                                                                              & \begin{tabular}[c]{@{}l@{}}Buypass Class 3 Root CA\\ Buypass AS-983163327, Norway\end{tabular}                                        \\ \cline{2-2} 
                                                                                                              & \begin{tabular}[c]{@{}l@{}}TWCA Root Certification Authority\\ TAIWAN-CA, Taiwan\end{tabular}                                         \\ \cline{2-2} 
                                                                                                              & \begin{tabular}[c]{@{}l@{}}Certum Trusted Network CA\\ Unizeto Technologies S.A., \\Certum Certification Authority, Poland\end{tabular} \\ \cline{2-2} 
                                                                                                              & \begin{tabular}[c]{@{}l@{}}T-TeleSec GlobalRoot Class 3\\T-Systems Enterprise Services GmbH, \\Germany\end{tabular}                    \\ \cline{2-2} 
                                                                                                              & \begin{tabular}[c]{@{}l@{}}Entrust Certification Authority - L1E \\Entrust, Inc., United States\end{tabular}                          \\ \hline
\end{tabular}
\end{center}
\end{table}
}{}

\appendix

\section{Survey questions}
Our survey contained seven questions (plus some log-specific questions based on our findings about the operators logs).  These questions are listed next:

\begin{itemize}
\item Do you setup the lists of acceptable roots for your logs manually or automatically?

\item How is the set of acceptable roots formed/updated?  (I.e., what is your policy for managing the list of acceptable root certificates?)

\item How often do you update your list(s) of acceptable roots?

\item Are there any procedures for root store verification and provision in place?

\item (If applicable) Some logs do not explicitly ship cross-origin (CORS) headers with their responses, which prevents browsers from querying the logs using javascript. Is this intentional? And if so, what is the reason?

\item In general, do you consider the process of root management transparent and secure enough?

\item (If applicable) What kind of improvements would you suggest to increase the security and transparency of the logs and to avoid future misconfigurations?

\end{itemize}

\section{Supporting details}

Tables~\ref{tab:duplicates} and~\ref{tab:duplicates2} list the specific roots that were duplicated in the JSON root lists of some of the logs, and the logs that contained these duplicates, as observed on Dec. 27, 2018, and Oct. 8, 2019, respectively.

\begin{table}[b]
 \centering
 \caption{Logs with duplicates in root lists (December 27, 2018)}
 \label{tab:duplicates}
\vspace{-8pt}
{\small
\begin{tabular}{ll}
\hline
\textbf{Logs}                                                                                                 & \textbf{Duplicated Root}                                                                                                                                                                                                         \\ \hline
\begin{tabular}[c]{@{}l@{}}DigiCert Log Server 2\\ DigiCert Yeti20* \\ DigiCert Nessie20* \end{tabular} & Atos TrustedRoot 2011                                                                                                                                                                                                            \\ \hline
\begin{tabular}[c]{@{}l@{}}GDCA Log 1\\ GDCA Log 2\end{tabular}                                               & \begin{tabular}[c]{@{}l@{}}DigiCert Trusted Root G4\\ Merge Delay Monitor Root\end{tabular}                                                                                                                                      \\ \hline
Venafi Gen2 CT log                                                                                            & \begin{tabular}[c]{@{}l@{}}Buypass Class 2 Root CA\\ Buypass Class 3 Root CA\\ TWCA Root Certification Authority\\ Certum Trusted Network CA\\ T-TeleSec GlobalRoot Class 3\\ Entrust Certification Authority - L1E\end{tabular} \\ \hline
\end{tabular}}
\end{table}

\begin{table}[b]
 \centering
 \caption{Logs with duplicates in root lists (October 8, 2019)}
 \label{tab:duplicates2}
\vspace{-8pt}
{\small
\begin{tabular}{ll}
\hline
\textbf{Logs}                                                                                                 & \textbf{Duplicated Root}                                                                                                                                                                                                         \\ \hline
\begin{tabular}[c]{@{}l@{}}DigiCert Log Server\end{tabular} & Atos TrustedRoot 2011 (1)                                                                                                                                                                                                       \\ \hline
\begin{tabular}[c]{@{}l@{}}DigiCert Log Server 2\end{tabular}                                               & \begin{tabular}[c]{@{}l@{}}Atos TrustedRoot 2011 (2)  \end{tabular}                     
\\ \hline
\begin{tabular}[c]{@{}l@{}}DigiCert Yeti20* \\ DigiCert Nessie20* \end{tabular}
      & \begin{tabular}[c]{@{}l@{}}Atos TrustedRoot 2011 (2) \\ VRK Gov. Root CA \end{tabular}  \\ \hline
\end{tabular}}
\vspace{-6pt}
\end{table}

\end{document}